\documentclass[aps,twocolumn,prl,floatfix,longbibliography,superscriptaddress,amssymb]{revtex4-1}
\pdfoutput=1

\usepackage{amsmath}
\usepackage{amsfonts}
\usepackage{amsbsy}
\usepackage{xcolor}
\usepackage{soul}
\usepackage{graphicx}
\usepackage{txfonts}
\usepackage{braket}
\usepackage{bm}
\usepackage[normalem]{ulem}

\usepackage[unicode]{hyperref}
\hypersetup{
    unicode=true, % non-Latin characters in Acrobat's bookmarks
    a4paper=true,
    plainpages=false,
    colorlinks=true,% false: boxed links; true: colored links
    linkcolor=blue,% color of internal links
    citecolor=blue,% color of links to bibliography
    filecolor=blue,% color of file links
    urlcolor=blue% color of external links
}
\newcommand{\pgej}{\rho_{ge}^{(j)}}
\newcommand{\peej}{\rho_{ee}^{(j)}}
\newcommand{\pgel}{\rho_{ge}^{(\ell)}}

\newcommand{\ddt}{\frac{\text{d}}{\text{d}t}}
\newcommand{\Isat}{I_{\text{sat}}}

\newcommand{\rj}{\mathbf{r}_j}

\newcommand{\omegaL}{\omega}

\newcommand{\psihat}[1]{\hat{\psi}_{#1}}
\newcommand{\psihatdag}[1]{\hat{\psi}_{#1}^{\dag}}
\newcommand{\vecHat}[2]{\hat{\mathbf{#1}}_{\mathrm{#2}}}

\newcommand{\cbEL}{\boldsymbol{\mathbf{\cal E}}}

\newcommand{\SC}{\textrm{SCEs}\xspace}
\newcommand{\QME}{\textrm{QME}\xspace}
\newcommand{\rjrN}{{\{\mathbf{r}_1,\dots,\mathbf{r}_N\}}}
\newcommand{\rhoN}{\rho_{\rjrN}}

\newcommand{\TcohQM}{T_{\mathrm{coh}}^{\mathrm{QM}}}
\newcommand{\TcohSC}{T_{\mathrm{coh}}^{\mathrm{SC}}}
\newcommand{\TincQM}{T_{\mathrm{inc}}^{\mathrm{QM}}}
\newcommand{\TincSC}{T_{\mathrm{inc}}^{\mathrm{SC}}}
\newcommand{\TincSAQ}{T_{\mathrm{inc}}^{\mathrm{SAQ}}}
\newcommand{\ODcohQM}{\mathrm{OD}_{\mathrm{coh}}^{\mathrm{QM}}}
\newcommand{\ODcohSC}{\mathrm{OD}_{\mathrm{coh}}^{\mathrm{SC}}}

\usepackage{xspace} % adds space after macros - alternatively use \QME{}

\begin{document}

\title{Quantum and Nonlinear Effects in Light Transmitted through Planar Atomic Arrays}
\author{Robert J. Bettles}
\affiliation{Joint Quantum Center (JQC) Durham--Newcastle, Department of Physics, Durham University, Durham DH1 3LE, UK}
\affiliation{ICFO-Institut de Ciencies Fotoniques, Mediterranean Technology Park, 08860 Castelldefels (Barcelona), Spain}
\author{Mark D. Lee}
\affiliation{Insight Risk Consulting, 16--18 Monument Street, Prospect Business Centres, London EC3R 8AJ, UK}
\author{Simon A. Gardiner}
\affiliation{Joint Quantum Center (JQC) Durham--Newcastle, Department of Physics, Durham University, Durham DH1 3LE, UK}
\author{Janne Ruostekoski}
\affiliation{Physics Department, Lancaster University, Lancaster LA1 4YB, UK} 
\date{\today}

\begin{abstract}
We identify significant quantum many-body effects, robust to position fluctuations and strong dipole--dipole interactions, in the forward light scattering from planar arrays and uniform-density disks of cold atoms, by comparing stochastic electrodynamics simulations of a quantum master equation and of a semiclassical model that neglects quantum fluctuations. 
Quantum effects are pronounced at high atomic densities, light close to saturation intensity, and around subradiant resonances. 
We show that such conditions also maximize  spin--spin correlations and entanglement of formation for the atoms, revealing the microscopic origin of light-induced quantum effects.
In several regimes of interest, an enhanced semiclassical model with a single-atom quantum description reproduces light transmission remarkably well, and permits analysis of otherwise numerically inaccessible large ensembles, in which we observe collective many-body analogues of resonance power broadening, vacuum Rabi splitting, and significant suppression in cooperative reflection from atomic arrays.
\end{abstract}

\maketitle

%% start main text %%
Light can mediate strong interactions between atoms, inducing strong position-dependent correlations, even in the limit of low light intensity, when the response (for the case of a simple level structure) is entirely classical. Such a correlated optical response can differ dramatically from that predicted by standard electrodynamics of continuous media, where resonant-light-induced dipole--dipole (DD) interactions between  atoms are treated in an averaged sense \cite{Javanainen2014shift,Javanainen2014a}. Beyond the limit of low light intensity, an isolated atom can scatter light quantum-mechanically, and quantum effects in the interactions of light with dilute atomic ensembles have been utilized in, e.g., quantum information protocols \cite{Hammerer2010}. In strongly interacting dense systems the possible role of quantum and cooperative effects is less clear and has been the subject of long-standing debates \cite{Javanainen2014shift, Friedberg1973, SCU10,ROH10,Guerin16b}. A particularly promising system to explore and utilize strong light-induced DD interactions is a regular planar array of scatterers such as atoms. In the linear low-excitation limit these manifest, as shown both theoretically and experimentally, a wealth of phenomena, e.g., subdiffraction features \cite{Sentenac2008,lemoult2010,Jenkins2012}, nontrivial topological phases \cite{Perczel2017a,Bettles2017}, transmission varying from complete reflection to full transparency \cite{Tretyakov,GarciadeAbajo2007,Jenkins2013,Bettles2016,Facchinetti2016,Facchinetti2018,Shahmoon2017}, narrow resonances and subradiance \cite{Fedotov2010,Jenkins2013,valentine2014,Facchinetti2016,Bettles2015d,Bettles2016,Jenkins2016b,Plankensteiner2017,Asenjo-Garcia2017a,Jen17,Guimond2019}, as well as quantum technological applications \cite{Ritsch_subr,Grankin18} and other collective effects \cite{Olmos13,Kramer2016,Yoo2016,Wang2017,Yoo18,Imamoglu2018b,Mkhitaryan2018}.

We show that we can identify quantum effects in the light transmitted through planar arrays and uniform-density disks of cold and dense atomic ensembles. 
Many-body quantum correlations are induced by light-atom coupling, which,
surprisingly, survive even strong many-body resonant DD interactions and atomic position fluctuations. Specifically, comparing the correlated optical response determined using the quantum master equation (\QME) to simulations neglecting any quantum fluctuations between atomic levels in different atoms [referred to as the ``semiclassical'' equations (\SC)~\cite{Lee2016}], we systematically identify light-established quantum effects between atoms in the transmitted light as a function of atom confinement, density, and driving intensity.
The effect of many-body quantum fluctuations on the scattering manifests most prominently at high densities when the light is close to saturation intensity, and especially significantly in the vicinity of subradiant resonances.
We find that these conditions also produce maximal spin--spin correlations and entanglement of formation in the underlying atomic system, further confirming the role of many-body quantum correlations and entanglement in observing a difference in light transmission between the \QME and \SC models. 
Incorporating the single-atom quantum description of light emission into the semiclassical scattering, we can typically use \SC also for incoherent scattering  to qualitatively reproduce the full quantum scattering even in the regimes where quantum effects in coherent
scattering were most pronounced, and elsewhere also quantitatively. 
 \SC therefore allow us to analyze cooperative transmission of light through large atomic arrays and disks beyond the limit of low light intensity, without needing to solve the full strongly-interacting quantum dynamics. Doing so, we find collective phenomena due to DD interactions that are many-body analogues of power broadening and vacuum Rabi splitting of atomic resonances in cavities \cite{Liberal2018,Khitrova2006}, and demonstrate a significant effect of intensity on the transmission
that may ultimately restrict the utilization of atomic arrays as highly reflective cooperative mirrors.

An appealing feature of light scattering from cold atoms~\cite{BalikEtAl2013,CHA14,Pellegrino2014a,Havey_jmo14,wilkowski,Jennewein2016,wilkowski2,Ye2016,Jenkins_thermshift,Guerin_subr16,Saint-Jalm2018} is that light-mediated strong DD interactions can establish correlations between atoms at fluctuating positions, which are most simply described using atomic field operators for the ground and excited states $\psihat{g,e}(\mathbf{r})$. Hence, $\langle\hat{\mathbf{P}}^+(\mathbf{r})\rangle=\langle \mathop{\psihatdag{g}(\mathbf{r})} \mathbf{d}_{ge} \mathop{\psihat{e}(\mathbf{r})}\rangle$ denotes  the light-induced atomic polarization, where $\mathbf{d}_{ge} = \mathbf{d}_{eg}^*$ is the dipole matrix element, and the populations are $\langle \mathop{\psihatdag{g}(\mathbf{r})}\mathop{\psihat{g}(\mathbf{r})}\rangle$ and $\langle \mathop{\psihatdag{e}(\mathbf{r})}\mathop{\psihat{e}(\mathbf{r})}\rangle$.
Because of the DD interactions, the polarization and populations also depend on two-body correlations $\langle \mathop{\psihatdag{a}(\mathbf{r})} \mathop{\psihatdag{b}(\mathbf{r}')} \mathop{\psihat{c}(\mathbf{r}')} \mathop{\psihat{d}(\mathbf{r})}\rangle$, where $a,b,c,d \in \{g,e\}$, representing the correlations in the optical response of an atom at $\mathbf{r}$ given the presence of a second atom at $\mathbf{r}'$. These in turn depend on three-body correlations, etc., resulting in a hierarchy of correlation function equations of motion \cite{Ruostekoski1997a,Morice1995a}. In a cold, dense ensemble this hierarchy can significantly and nonperturbatively modify the scattering behavior, invalidating attempts to truncate it \cite{Javanainen2014shift}. This is a key ingredient in, e.g., Anderson localization of light, which has been a subject of considerable controversy and debate \cite{Skipetrov2014,Sperling2016}.

A numerical device for solving this correlation function hierarchy is to treat the atoms as discrete point particles, meaning for a particular configuration of atomic positions $\{\mathbf{r}_1,\dots,\mathbf{r}_N\}$ that two-body correlation functions take the form 
\begin{multline}
\left\langle \mathop{\psihatdag{a}(\mathbf{r})} \mathop{\psihatdag{b}(\mathbf{r}')} \mathop{\psihat{c}(\mathbf{r}')} \mathop{\psihat{d}(\mathbf{r})}\right\rangle_{\{\mathbf{r}_1,\dots,\mathbf{r}_N\}} 
\\
= \sum_{ j\ell (j\neq\ell)
} \rho_{ad;bc}^{(j,\ell)}\mathop{\delta(\mathbf{r}-\mathbf{r}_j)} \mathop{\delta(\mathbf{r}'-\mathbf{r}_{\ell})},
\label{eq:twobody}
\end{multline}
where $\rho_{ad;bc}^{(j,\ell)}$ denote correlation functions of the internal atomic energy levels only \cite{Lee2016}. We then solve the internal atom dynamics at discrete positions, and the new correlation functions simply emerge from the $N$-body density matrix $\rho \equiv \rhoN$. This evolves according to \QME 
\begin{align} \label{eq:MasterEquation}
\dot \rho =&  -\frac{\text{i}}{\hbar}\left[\sum_j\mathsf{H}_{\text{sys},j} - \sum_{j\ell(j\neq\ell)} \hbar\Omega_{j\ell} \hat{\sigma}_+^{(j)} \hat{\sigma}_-^{(\ell)} , \rho \right] \nonumber \\
 & + \sum_{j\ell}\gamma_{j\ell}\left( 2 \hat{\sigma}_-^{(j)} \rho \,\hat{\sigma}_+^{(\ell)} - \hat{\sigma}_+^{(\ell)} \hat{\sigma}_-^{(j)}\rho - \rho \hat{\sigma}_+^{(\ell)}\hat{\sigma}_-^{(j)} \right),
\end{align}
where the collective scattering  is represented by the dispersive $\Omega_{j\ell}$ and dissipative $\gamma_{j\ell}$ DD interactions, the single-atom half-width at half-maximum (HWHM) linewidth by $\gamma_{jj}=\gamma$, and 
$\hat{\sigma}_+^{(j)} = (\hat{\sigma}_-^{(j)})^{\dag} = 
|e\rangle_{j}\mbox{}_{j}\langle g|$. 
 For simplicity, we consider two-level atoms and the Hamiltonian 
\begin{align} \label{eq:Hsys}
\mathsf{H}_{\text{sys},j} = -\hbar\Delta\hat{\sigma}^{(j)}_{ee} -
\mathbf{d}_{eg} \cdot \mathop{{\cbEL}^+(\mathbf{r}_j)}  
\hat{\sigma}^{(j)}_{+} 
-\mathbf{d}_{ge}  \cdot \mathop{{\cbEL}^-(\mathbf{r}_j)} 
\hat{\sigma}^{(j)}_{-} ,
\end{align}
where $\cbEL^+ = (\cbEL^-)^*$ is the positive-frequency-component of the frequency $\omegaL$ laser field, detuned from the atomic resonance frequency $\omega_{ge}$ by $\Delta=\omegaL-\omega_{ge}$ and $\hat{\sigma}_{ee}^{(j)} = |e\rangle_{j}\mbox{}_{j}\langle e|$.
Spatial correlations are numerically synthesized by ensemble-averaging over stochastic realizations of atomic positions sampled from the density distribution \cite{Javanainen1999,Lee2016}.
Solving Eq.~(\ref{eq:MasterEquation}) for large systems is numerically taxing, although few-atom ensembles already demonstrate many-body effects in their spectra \cite{Jones2017}.

In the limit of low light intensity, where the excited state population vanishes, the internal level correlations, such as those described by $\rho_{ad;bc}^{(j,\ell)}$ in Eq.~(\ref{eq:twobody}), also vanish for two-level atoms. The stochastic electrodynamics simulations are then formally exact \cite{Lee2016,Javanainen1999}, reproducing the many-atom spatial correlations, which are identical to those occurring in the classical electrodynamics of coupled linear electric dipoles.
Beyond the limit of low light intensity, the full dynamics of Eq.~(\ref{eq:MasterEquation}) can be greatly simplified by factorizing the internal atomic level correlation functions:
\begin{equation} 
\rho_{ad;bc}^{(j,\ell)} \simeq \rho_{ad}^{(j)}\rho_{bc}^{(\ell)}.
\label{eq:fact}
\end{equation}
Following the formalism of 
\cite{Lee2016} 
we then  obtain coupled nonlinear equations
\begin{align}
\label{eq:pgeEOM}
\ddt \pgej = & (\text{i}\Delta-\gamma)\pgej - \frac{\rm{i}}{\hbar}\left(2\peej-1\right) \mathbf{d}_{eg}\cdot \mathop{\cbEL^+(\rj)} \nonumber \\
& - \text{i}\left(2\peej-1\right)\sum_{\ell\neq j}\left( \Omega_{j\ell} + \rm{i}
\gamma_{j\ell} \right)\pgel,
\\
\label{eq:peeEOM}
\ddt \peej = & -2\gamma \,\peej + \frac{2}{\hbar} \text{Im}\left[ \mathop{\cbEL^-(\rj)} \cdot \mathbf{d}_{ge}\,\pgej\right] \nonumber \\
& + 2 \,\text{Im} \left[ \sum_{\ell\neq j} \pgej \left( \Omega_{j\ell} - \rm{i}\gamma_{j\ell} \right) {\pgel}^* \right].
\end{align}
Note the relatively small number of equations $2N$ compared to the full quantum system size $2^N$.
This formalism has been applied to the modeling of pumping of atoms in dense clouds \cite{Machluf2018}, and has also been extended to cavity QED \cite{Lee2017}.

%% figure %%
\begin{figure}[tb]
  \centering
	\includegraphics[width=0.8\columnwidth]{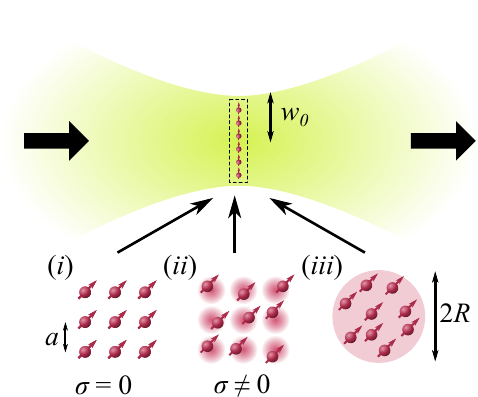}	
	\caption{Schematic illustration of  a laser beam (with waist $w_0$) focused onto an ensemble of atoms arranged in ($i$,$ii$) a planar square array with the spacing $a$ and  ($i$) fixed positions;  ($ii$) normally-distributed position fluctuations with standard deviation $\sigma$; ($iii$) random positions on a disk of uniform density and radius $R$. 
	}
	\label{fig:Schematic}
\end{figure}
%%

%% figure %%
\begin{figure*}[tb]
    \centering
    \includegraphics[width=2\columnwidth]{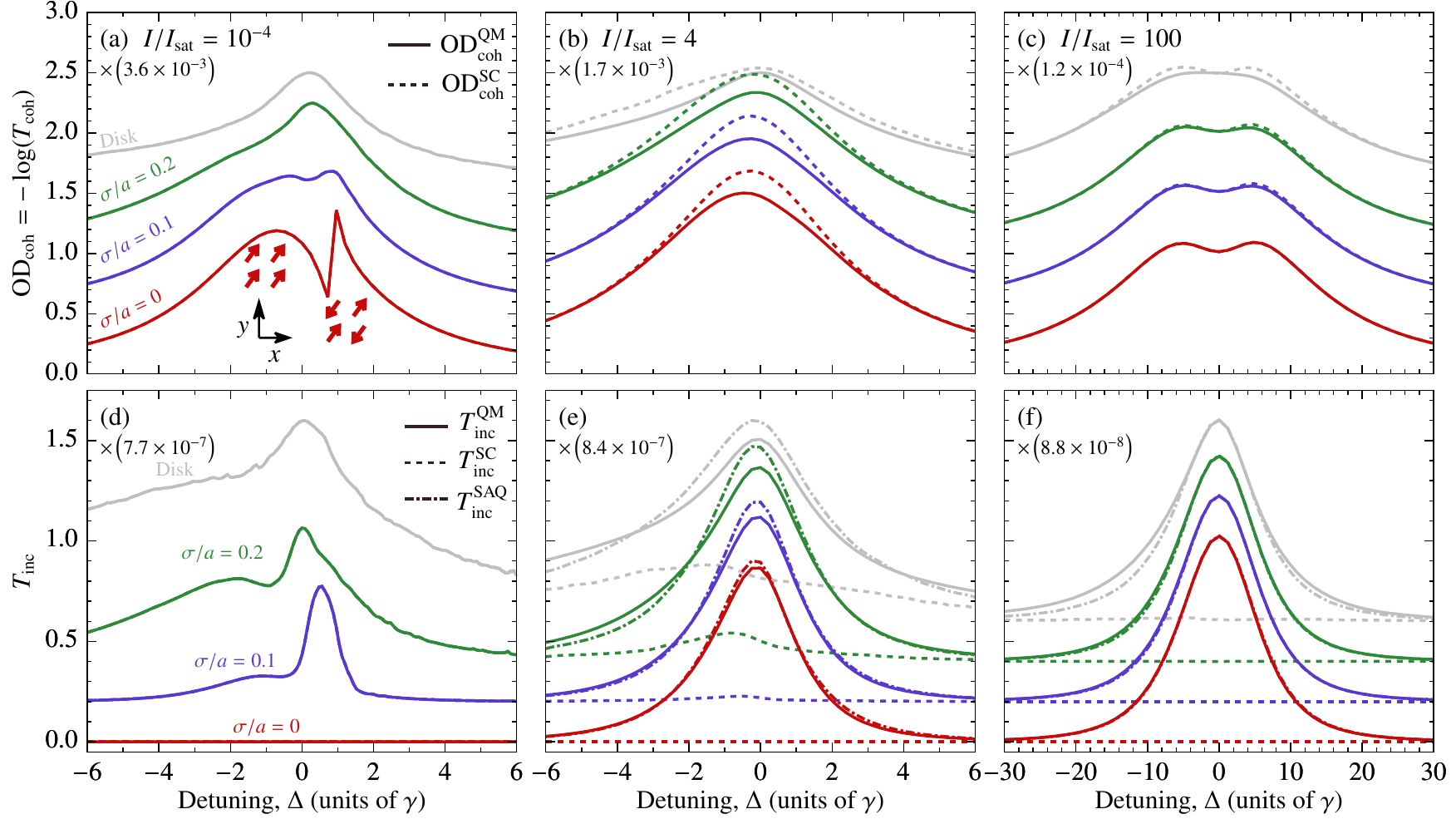}
	\caption{
	Quantum effects in the optical transmission through a strongly coupled atomic ensemble. Lines from bottom to top: atoms trapped in a $2\times 2$ square array in the $xy$ plane with lattice constant $a=0.25\lambda$ and position fluctuations determined by normal distributions with standard deviation $\sigma_{x,y}/a = 0, 0.1, 0.2$, respectively, and the atoms uniformly distributed inside a disk of radius $R=0.28\lambda$. Except for the case of fixed atomic positions,  $\sigma_z=0.025\lambda$. The peak disk density is $1.0 k^{3}$. For clarity each line is offset from the one below it by 0.5 (a--c) and 0.2 (d--f). 
	The incident field has beam waist $w_0=10\lambda$, polarization $(\hat{\mathbf{x}} + \hat{\mathbf{y}})/\sqrt{2}$, and intensity $I/\Isat=0.0001$ (a,d), $I/\Isat=4$ (b,e), and $I/\Isat=100$ (c,f). 
	Red arrows in (a) show the dipole vectors in the $xy$ plane for the super-radiant (left) and subradiant (right) eigenvectors.
}
    \label{fig:CoherentIncoherentScattering}
\end{figure*}
%%

%% figure %%
\begin{figure}[tb]
    \centering
    \includegraphics[width=\columnwidth]{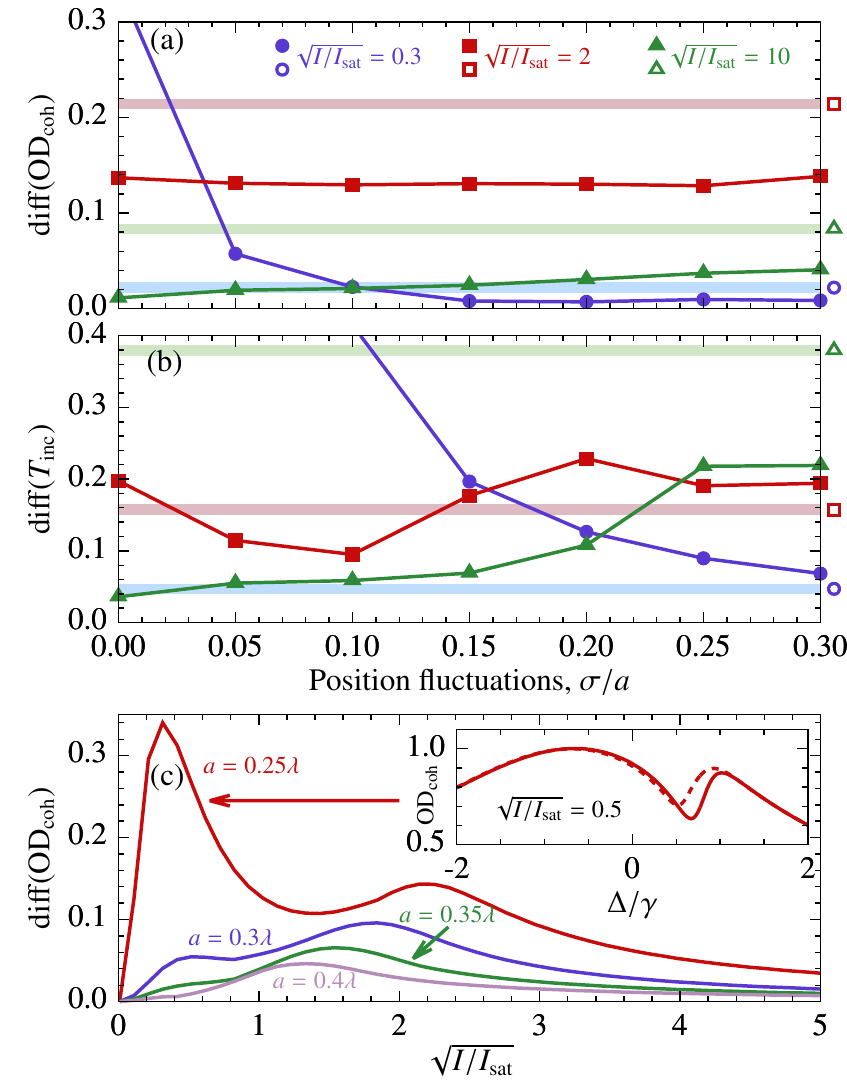}
        \caption{Quantum many-body effects via the relative difference between $\ODcohQM=-\log(\TcohQM)$ and $\ODcohSC$ (a,c), and $\TincQM$ and $\TincSAQ$ (b), for different intensities and position fluctuations (a,b) and lattice spacings (c). 
        The parameters are the same as in Fig.~\ref{fig:CoherentIncoherentScattering} and $\mathrm{diff}(S) \equiv \max \{|S_{\SC}(\Delta)-S_{\QME}(\Delta)|/S_{\QME}(\Delta)\} $, for $\Delta/\gamma\in[-4,4]$ (a,b) and $\Delta\in[-20,20]$ (c).
        (a,b): arrays (solid markers), disk traps (open markers); The inset: $\ODcohQM$ (solid) and $\ODcohSC$ (dashed) for $a=0.25\lambda$ and $\sqrt{I/\Isat}=0.5$.
        }
    \label{fig:ComparingModels}
\end{figure}
%%

%% figure %%
\begin{figure}[tb]
  \centering
	\includegraphics[width=\columnwidth]{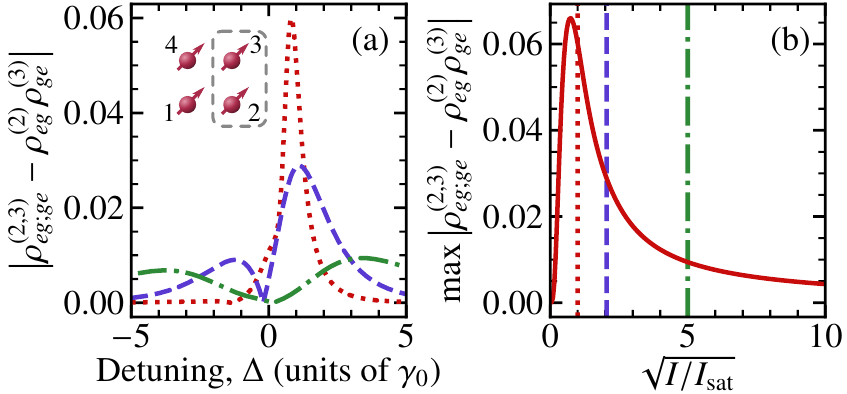}
	\caption{Spin--spin correlations showing the emergence of light-induced quantum correlations and deviations from the semiclassical approximation of Eq.~\eqref{eq:fact}, for the same $2\times 2$ array as in Figs.~\ref{fig:CoherentIncoherentScattering},\ref{fig:ComparingModels}. For fixed atomic positions the origin of the correlations is solely quantum, and they appear between internal levels of different atoms. (a) The correlation lineshape for increasing intensity ($\sqrt{I/\Isat}=1$ -- red dotted line, $2$ -- blue dashed line, $5$ -- green dot--dashed line) exhibits an energy splitting. 
	(b) The correlations peak around $I\sim\Isat$, going to zero for very low and very high light intensities.
	The behavior is qualitatively similar for other spin and atom pairs in the system. 
	}
	\label{fig:Correlations}
\end{figure}
%%

%% figure %%
\begin{figure}[tb]
  \centering
	\includegraphics[width=\columnwidth]{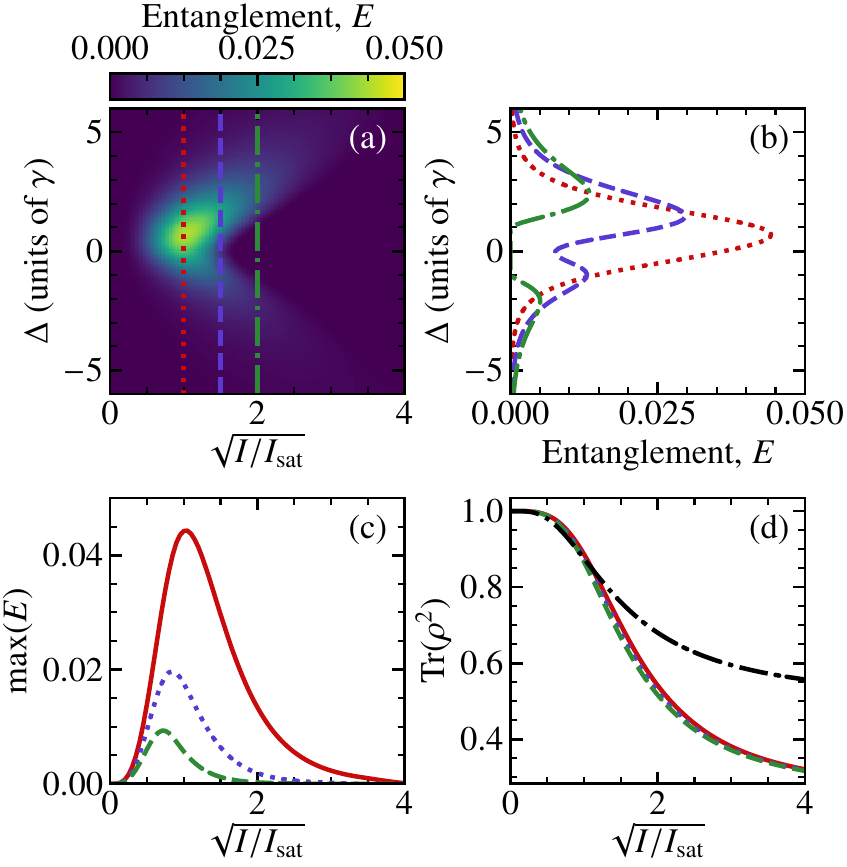}
	\caption{
	Entanglement of formation $E$ \cite{Wootters1998} (a,b,c) and trace purity (d) for a pair of two-level atoms separated in $y$, polarized in $y$, and driven uniformly by an incident field with intensity $I$. 
	The optical response of this two-atom system is analogous to that of four atoms in Figs.~\ref{fig:CoherentIncoherentScattering}-\ref{fig:Correlations}.
	The atomic spacing is (a,b) $a=0.25\lambda$; (c,d) $0.25\lambda$ (red solid lines), $0.3\lambda$ (blue dotted lines), and $0.35\lambda$ (green dashed lines). (a,b) Entanglement of formation as a function of detuning for (a) varying light intensity; (b) $\sqrt{I/\Isat}=1$ (red dotted lines), $1.5$ (blue dashed lines), and $2$ (green dot--dashed lines). At higher intensities $I>I_{\mathrm{sat}}$, the entanglement resonance splits (as the correlations in Fig.~\ref{fig:Correlations}) and $E$ is no longer maximized at the atomic resonance. (c) The maximum entanglement decreases for increasing atomic spacing and peaks around $I=I_{\mathrm{sat}}$, decreasing to 0 as $\sqrt{I/I_{\mathrm{sat}}}\to0$ or $\infty$. (d) The purity of the atomic state decreases towards the limit of $1/4$ for increasing $I$ and  is less sensitive to the array spacing than $E$. For reference, the purity for a single atom is shown in black (dot--dashed).
	}
	\label{fig:Entanglement}
\end{figure}
%%

%% figure %%
\begin{figure}[tb]
    \centering
    \includegraphics[width=\columnwidth]{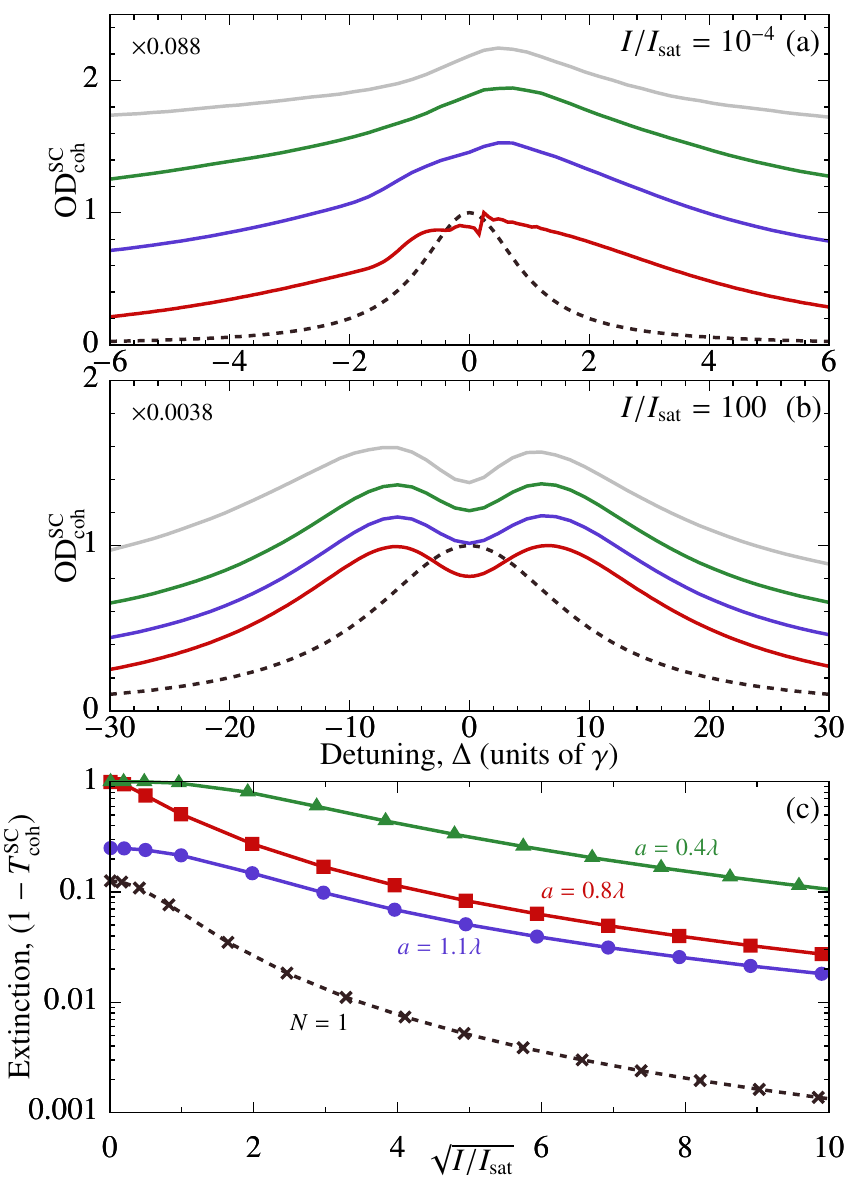}
    \caption{Semiclassically evaluated optical depth (a,b) and maximum extinction (c) for $N=100$ atoms. (a,b) Parameters as in Fig.~\ref{fig:CoherentIncoherentScattering} except fusing a $10\times 10$ square array (bottom three lines) and a disk with radius $R=1.4\lambda$ and peak density $1.0 k^{3}$ (top lines). Black dashed lines show the lineshape for a single atom with power-broadened $\gamma_{\mathrm{PB}}=\gamma(1+I/\Isat)^{1/2}$. (c) Peak extinction ($1-\TcohSC$) through a $10\times 10$ fixed lattice with constant $a=0.8\lambda$ and beam waist $w_0=3\lambda$ (squares), $a=1.1\lambda$ and $w_0=3\lambda$ (circles), and $a=0.4\lambda$ and $w_0=1.5\lambda$ (triangles). In (c), the laser and atomic polarization is $-(\hat{\mathbf{x}}+\mathrm{i}\hat{\mathbf{y}})/\sqrt{2}$ and the light collected through the scattering angle $\sin\theta\alt 0.37$.
    }
    \label{fig:SquareLatticeLineshapes}
\end{figure}

Spatially correlated scattering between different atoms is accounted for in Eqs.~(\ref{eq:pgeEOM}) and~(\ref{eq:peeEOM}) via $\Omega_{j\ell}$ and $\gamma_{j\ell}$ (for $\Omega_{j\neq\ell}=\gamma_{j\neq\ell}=0$ they reduce to the independent-atom optical Bloch equations).
In the limit of low light intensity the ensemble-averaged response of \SC coincides with the exact classical electrodynamics; beyond this limit, the model incorporates nonlinear internal level dynamics of the atoms.
However, because of the factorization in Eq.~\eqref{eq:fact}, they cannot account for many-body quantum entanglement between different atoms' internal levels.
Finding situations in which the predictions of \SC observably differ from the full \QME solution therefore identifies light-induced quantum effects in the transmitted light. Conversely, regimes where quantum fluctuations are minimal allow for the simulation of much larger systems than are accessible with \QME, and also test the validity of related approaches in other contexts, based, e.g., on mean-field approximations, intensity expansions, or truncations of the correlations \cite{Kramer2015a,Sutherland_satur,Zhang2018,Henriet2018,Parmee2018}.

We begin by calculating the coherent and incoherent forward transmission, $T_{\mathrm{coh}}$ and $T_{\mathrm{inc}}$  (Figs.~\ref{fig:CoherentIncoherentScattering} and \ref{fig:ComparingModels}) \cite{SuppMat}, through planar square arrays and thin disks of $N=4$ atoms (Fig.~\ref{fig:Schematic}). The array could be realized, e.g., by an optical lattice \cite{Gross2017} or dipole traps \cite{Nogrette2014,Bernien2017}. Unless otherwise stated, we consider lattice spacing $a=0.25\lambda$ and disk radius $R=0.28\lambda$. Physically, we calculate the far-field light intensity in the same mode as the driving field $\cbEL^{\pm}$, integrated over the polar angle $\sin\theta\alt 0.24$ \cite{SuppMat}.
We account for the fluctuations in atomic positions due to finite trap confinement by ensemble-averaging over many stochastic realizations of position configurations \cite{Jenkins2012,SuppMat}. We unambiguously identify quantum effects in coherent transmission from the difference between the full quantum and semiclassical transmissions $\TcohQM - \TcohSC$. Since the coherent scattering of a single atom is classical, this difference is due solely to many-body quantum correlations in the atomic response. 

To obtain the incoherently scattered light $\langle \delta\vecHat{E}{d}^-(\mathbf{r}) \,\delta\vecHat{E}{d}^+(\mathbf{r}) \rangle$, we write the scattered light field as  $\vecHat{E}{d}^+ = \langle \vecHat{E}{d}^+ \rangle + \delta \vecHat{E}{d}^+$, where $\delta\vecHat{E}{d}^+$ denotes the fluctuations \cite{Meystre2007}. This yields incoherent transmission \cite{SuppMat} for which quantum behavior also is isolated by $\TincQM-\TincSC$. We can improve the semiclassical incoherent model, without increasing the computational complexity, by adding the single-atom quantum description of incoherent light emission for all the atoms. In a single realization of stochastic atomic positions, the incoherent scattering contribution to intensity from independent quantum-mechanical atoms  $\propto \sum_j A_j( \langle \hat{\sigma}^{(j)}_{ee} \rangle -| \langle \hat{\sigma}^{(j)}_+\rangle |^2  )$, where $A_j$ encapsulates the light propagation effects \cite{Meystre2007,SuppMat}. Augmenting the semiclassical model with this single-atom quantum description integrated over the sample yields the incoherent transmission $\TincSAQ$. The many-body quantum effects of the incoherent signal are then encapsulated in $\TincQM-\TincSAQ$.

In Fig.~\ref{fig:CoherentIncoherentScattering}(b), we identify many-body quantum fluctuations in the coherent transmission ($\TcohQM-\TcohSC$) that increase with increasing DD interaction [Fig.~\ref{fig:ComparingModels}(a,c)], reaching normalized residuals of over $10\%$ at $a=0.25\lambda$ and $I\simeq \Isat$ (when the dipole amplitudes are greatest), where 
${I=2\epsilon_0 c|\cbEL^+(\mathbf{r}=\mathbf{0})|^2}$
is the total intensity of the incident laser field at the focus and $\Isat = 4\pi^2\hbar\gamma c/3\lambda^3$ is the saturation intensity. Strikingly, quantum effects constitute over $30\%$ of the signal in the vicinity of the narrow subradiant resonant shown in the inset of Fig.~\ref{fig:ComparingModels}(c). This may be due to the enhanced dipole magnitude of subradiance \cite{Bettles2016}, the antisymmetry of the collective dipolar eigenvector [see diagram in Fig.~\ref{fig:CoherentIncoherentScattering}(a)], or the rapidly varying Fano interference.
It is remarkable that even for a fully random disk quantum effects on the scattering are not washed out but can produce residuals between the models of a few percent.

Conversely, as in Fig.~\ref{fig:CoherentIncoherentScattering}(e,f), once $I\agt \Isat$ the incoherent transmission is almost entirely dominated by quantum fluctuations ($\TincSC\to 0$ \footnote{See also Fig.~S1 of \cite{SuppMat}}). However, once we incorporate the single-atom quantum description into the scattering and therefore transmission $\TincSAQ$, the difference becomes much smaller and the many-body quantum fluctuations are, as with the coherent scattering, maximal around $I\sim\Isat$. Hence, using the improved model $\TincSAQ$, it is possible, even for incoherent scattering,  to obtain excellent qualitative, and frequently quantitative agreement with the full quantum scattering. For example, in Fig.~\ref{fig:ComparingModels}(c) the difference between $\TcohQM$ and $\TcohSC$ is less than $5\%$ for $a\gtrsim 0.4\lambda$ or $I/\Isat\gtrsim16$.

Up until now, we have identified many-body quantum effects in the transmitted light. These originate from the light-induced quantum correlations between internal levels of different atoms that do not satisfy the factorization assumption of the \SC, given in Eq.~\eqref{eq:fact}. We explicitly show these induced spin--spin correlations and many-body entanglement of formation (in the latter case, for an analogous configuration of a pair of atoms, using the formalism of \cite{Wootters1998})  in Figs.~\ref{fig:Correlations} and \ref{fig:Entanglement}, respectively. The conditions under which the spin--spin correlations and the entanglement of formation are maximal do indeed match the conditions for enhanced quantum effects in the light transmission in Figs.~\ref{fig:CoherentIncoherentScattering} and~\ref{fig:ComparingModels}.

Working in the conditions in which the quantum effects on the light scattering are minimal,  we can employ
\SC [Eqs.~(\ref{eq:pgeEOM}) and (\ref{eq:peeEOM})] by neglecting quantum fluctuations to analyze the coherent transmission through much larger ensembles, for which the full \QME is inaccessible.
In Fig.~\ref{fig:SquareLatticeLineshapes} we show how the transmission lineshapes of a $10\times 10$ array significantly differs from the Lorentzians of independent atoms. For a single atom the linewidth is power broadened to $\gamma_{\mathrm{PB}}(I)=\gamma\sqrt{1+I/\Isat}$. In the interacting case, the coherent lineshape is also power broadened but, depending on whether the linewidth of the dominant symmetric collective eigenmode $\upsilon$ at low light intensity \cite{SuppMat} is subradiant ($\upsilon <\gamma$) or superradiant ($\upsilon >\gamma$), it will also be narrower or broader [Fig.~\ref{fig:SquareLatticeLineshapes}(b)], respectively, than $\gamma_{\mathrm{PB}}(I)$. There is no analogous broadening of 
the incoherent lineshape, 
however \footnote{See Fig.~S2 of \cite{SuppMat}}.
Furthermore, the many-body lineshape also exhibits a dip or ``hole burning'' on resonance [Fig.~\ref{fig:SquareLatticeLineshapes}(b)]. This dip is analogous to vacuum Rabi splitting \cite{Liberal2018,Khitrova2006}, where the interatomic DD coupling has now taken the role of the cavity coupling, and, while it only occurs for sufficiently high density, it can interestingly still exist even in the fully random ensemble. 
 
A key feature of general subwavelength-spaced resonant emitter arrays is that they can exhibit perfect reflection \cite{Tretyakov,GarciadeAbajo2007}, which may typically be modeled using point-dipole scatterers \cite{Jenkins2013,Shahmoon2017,Bettles2016,Facchinetti2016}.
Dipolar planar arrays can act as cooperative antennae \cite{AdamoEtAlPRL2012}, with applications to quantum information processing \cite{Grankin18}, making understanding nonlinear transmission essential.
We calculate this for large arrays in Fig.~\ref{fig:SquareLatticeLineshapes}(c), and find that the reduction in the extinction as a function of light intensity is considerable --- although less prominent with smaller spacings. This may ultimately restrict the applications of atomic arrays as highly reflective cooperative mirrors to weak light intensities only.

To conclude, by comparing \SC and \QME, we have identified light-induced spin--spin correlations and quantum entanglement in the light transmitted through planar arrays and disks which survive both position fluctuations and strong DD interactions. 
At narrow subradiant resonances, quantum fluctuations can be over $30\%$. 
Outside these resonances, provided we improve the model by incorporating the single-atom quantum description, \SC typically still reproduce, also for incoherent scattering, the full quantum behavior at least qualitatively. 
This provides a methodology to calculate transmission of light through large arrays, consisting of hundreds of atoms, which can exhibit striking many-body phenomena (even without any quantum effects) reminiscent of single-atom power broadening and vacuum Rabi splitting. 
The existence of many-body quantum effects despite strong driving, high densities, and even with significant atomic position fluctuations is surprising. It suggests that optical quantum information processing in atomic ensembles~\cite{Hammerer2010} need not necessarily be restricted to dilute systems.
Subradiant resonance narrowing has now been experimentally observed in the transmitted light through an optical lattice of atoms
in a Mott-insulator state in the classical limit of low light intensity \cite{Rui2020}. Several of our findings could also be verified in this setup by increasing the intensity of the incident light.
The presence, even in uniform disks, of many-body effects, that are attracting considerable interest \cite{Javanainen2014shift,Javanainen2014a,Saint-Jalm2018,Keaveney2012}, further relaxes the conditions necessary for their experimental observation.

%% acknowledgements %%
\begin{acknowledgements}
J.R.\ acknowledges financial support from the UK EPSRC (Grant Nos.\ EP/P026133/1, EP/M013294/1).  M.D.L.\ and J.R.\ also acknowledge support from the UK EPSRC (EP/H049568/1). R.J.B.\ acknowledges financial support from the CELLEX ICFO-MPQ Postdoctoral Fellowship program, the Spanish MINECO Severo Ochoa Grant No.\ SEV 2015-0522, CERCA Programme/Generalitat de Catalunya, and Fundacio Privada Cellex; R.J.B.\ with S.A.G.\ also acknowledge support from the UK EPSRC (EP/R002061/1). 
\end{acknowledgements}
%%

%% bibliography %%
\bibliography{References_arxiv}

\end{document}

\title{Supplemental Material to \\``Quantum and Nonlinear Effects in Light Transmitted through Planar Atomic Arrays''}
\author{Robert J. Bettles}
\affiliation{Joint Quantum Center (JQC) Durham--Newcastle, Department of Physics, Durham University, Durham DH1 3LE, UK}
\affiliation{ICFO-Institut de Ciencies Fotoniques, Mediterranean Technology Park, 08860 Castelldefels (Barcelona), Spain}
\author{Mark D. Lee}
\affiliation{Insight Risk Consulting, 16--18 Monument Street, Prospect Business Centres, London EC3R 8AJ, UK}
\author{Simon A. Gardiner}
\affiliation{Joint Quantum Center (JQC) Durham--Newcastle, Department of Physics, Durham University, Durham DH1 3LE, UK}
\author{Janne Ruostekoski}
\affiliation{Physics Department, Lancaster University, Lancaster LA1 4YB, UK} 
\date{\today}

\maketitle

%% custom supplemental style %%
\renewcommand{\thesection}{S\Roman{section}}
\renewcommand{\thesubsection}{S\Roman{section}.\Alph{subsection}}
\renewcommand{\theequation}{S\arabic{equation}}
\renewcommand{\thefigure}{S\arabic{figure}}
\renewcommand{\bibnumfmt}[1]{[S#1]}
\renewcommand{\citenumfont}[1]{S#1}

%% begin text %%
\section{Dynamics and correlation functions}

\noindent We simulate the optical response of $N$-atom ensembles by 
stochastically sampling fixed positions $\{\mathbf{r}_1,\dots,\mathbf{r}_N\}$ of stationary atoms, as the atomic center-of-mass dynamics are assumed negligible. In the full quantum dynamics, for each stochastic realization we solve the equations of motion for the $N$-atom density matrix $\rhoN (t)$ with the atoms at fixed positions $\{\mathbf{r}_1,\dots,\mathbf{r}_N\}$, obeying
the quantum master equation (\QME)  [Eq.~(2) in the main text],
\begin{widetext}
\begin{equation} \label{eq:MasterEquationSupp}
\ddt \rhoN =  -\frac{\text{i}}{\hbar}\left[\sum_j\mathsf{H}_{\text{sys},j} - \sum_{j\ell(j\neq\ell)} \hbar\Omega_{j\ell} 
\hat{\sigma}_{+}^{(j)} \hat{\sigma}_{-}^{(\ell)}, 
\rhoN \right]
 + \sum_{j\ell}\gamma_{j\ell}\left[ 2 \hat{\sigma}_{-}^{(j)} \rhoN \hat{\sigma}_{+}^{(\ell)}
 -
 \hat{\sigma}_{+}^{(\ell)} \hat{\sigma}_{-}^{(j)}\rhoN
 -
 \rhoN \hat{\sigma}_{+}^{(\ell)}\hat{\sigma}_{-}^{(j)} \right].
\end{equation}
\end{widetext}
The single atom Hamiltonian $\mathsf{H}_{\text{sys},j}$ (in which we have assumed the rotating wave approximation) has the form 
\begin{align}
\mathsf{H}_{\text{sys},j} = -\hbar\Delta\hat{\sigma}^{(j)}_{ee} -
\mathbf{d}_{eg} \cdot \mathop{{\cbEL}^+(\mathbf{r}_j)}  
\hat{\sigma}^{(j)}_{+} 
-\mathbf{d}_{ge}  \cdot \mathop{{\cbEL}^-(\mathbf{r}_j)} 
\hat{\sigma}^{(j)}_{-} ,
\end{align}
where $\Delta=\omegaL - \omega_{ge}$ is the detuning of the laser frequency $\omegaL$ from the atomic transition frequency $\omega_{ge}$, $\mathbf{d}_{ge} = \mathbf{d}_{eg}^*$ is the dipole matrix element, $ {\cbEL}^{+}$ is the positive frequency component of the laser amplitude  (given in terms of the electric displacement $\DL^+=\epsilon_0 \cbEL^+$), and the raising operator from the ground state $|g\rangle$ to the excited state $|e\rangle$,
$\hat{\sigma}_{+}^{(j)}=|e\rangle_{j}\mbox{}_{j}\langle g|$,
lowering operator 
$\hat{\sigma}_-^{(j)}=|g\rangle_{j}\mbox{}_{j}\langle e|$, 
and excited state population operator 
$\hat{\sigma}_{ee}^{(j)}=|e\rangle_{j}\mbox{}_{j}\langle e|= 1- |g\rangle_{j}\mbox{}_{j}\langle g|$
are single-atom operators for the $j$th atom. We use slowly-varying field amplitudes and atomic variables where the rapid rotation at the laser frequency $\omega$
has been factored out by substitutions $ {\cbEL}^{+} \mathrm{e}^{\mathrm{i}\omega t} \rightarrow {\cbEL}^{+}$, $ {\hat\sigma}^{(j)}_{+} (t) \mathrm{e}^{\mathrm{i}\omega t}\rightarrow {\hat\sigma}^{(j)}_{+} (t) $, etc. 
The collective coupling matrices $\Omega_{j\ell}$ and $\gamma_{j\ell}$, resulting, respectively, in collective resonance line shifts and linewidths in Eq.~\eqref{eq:MasterEquationSupp}, are the real and imaginary parts of the dipole radiation kernel $\mathsf{G}(\mathbf{r})$: 
\begin{equation} \label{eq:G_Omegagamma}
\frac{1}{\hbar \epsilon_0}\mathbf{d}_{eg}\cdot\left[\mathop{\mathsf{G}(\mathbf{r}_{j}-\mathbf{r}_{\ell})} 
\mathbf{d}_{ge}\right] = \Omega_{j\ell} + \mathrm{i}\gamma_{j\ell},
\end{equation}
where
\begin{widetext}
\begin{equation}  
{\sf G}({\bf r})\,\hat{\bf d}=
{k^3\over4\pi}
\left\{ (\hat{\bf n}\!\times\!\hat{\bf d}
)\!\times\!\hat{\bf n}{e^{ikr}\over kr}
+\big[3\hat{\bf n}(\hat{\bf n}\cdot\hat{\bf d})-\hat{\bf d}\big]
\big[ {1\over (kr)^3} - {i\over (kr)^2}\big]e^{ikr}
\right\}-{\hat{\bf d}\,\delta({\bf r})\over3}
 \label{eq:GreensFunction}
\end{equation}
\end{widetext}
is the electric field amplitude for an oscillating electric dipole at the origin, $\hat{\bf n} = {{\bf r}/ r}$, and $k=2\pi/\lambda$ for laser wavelength $\lambda$. Note that we typically drop the contact interaction term \cite{Lee2016}.

Once the full density matrix $\rhoN (t)$ for a particular set of fixed atomic positions $\{\mathbf{r}_1,\dots,\mathbf{r}_N\}$ is known, the 
one-body $\rho_{ab}^{(j)}$ ($j$th atom), two-body $\rho_{ad;bc}^{(j,\ell)}$  ($j$th and $\ell$th atoms), etc., expectation values for this stochastic realization are given by
\begin{align}
\rho_{ge}^{(j)} = & \left\langle \hat{\sigma}_+^{(j)}\right\rangle = \mathrm{Tr}\left\{\hat{\sigma}_+^{(j)}\rhoN\right\},
\\
\rho_{ee}^{(j)} =& 1 - \rho_{gg}^{(j)} = \left\langle\hat{\sigma}_{ee}^{(j)}\right\rangle = \mathrm{Tr}\left\{ \hat{\sigma}_{ee}^{(j)}\rhoN\right\}, 
\end{align}

\pagebreak
\noindent\begin{align}
\rho_{eg;ge}^{(j,\ell)} = & \left \langle \hat{\sigma}_-^{(j)} \hat{\sigma}_+^{(\ell)} \right \rangle (1-\delta_{j\ell})
=\mathrm{Tr}\left\{\hat{\sigma}_-^{(j)} \hat{\sigma}_+^{(\ell)} \rhoN \right\}(1-\delta_{j\ell}),
\end{align}
and so forth.

In each stochastic realization, the $N$-atom configuration of  positions $\{\mathbf{r}_1,\dots,\mathbf{r}_N\}$ is obtained by sampling from a joint probability distribution $P(\mathbf{r}_1,\dots,\mathbf{r}_N)$, taken to be the initial distribution of stationary atoms.
Ensemble-averaging over many such realizations then transforms the expectation values  $\rho_{ab}^{(j)}(t)$, $\rho_{ad;bc}^{(j,\ell)}(t)$, etc.,  to spatial correlation functions for the atoms 
at any given time $t$:
\begin{widetext}
\begin{align} 
\label{eq:sampling}
\left\langle 
\mathop{\hat{\psi}^{\dag}_a(\mathbf{r},t)} 
\mathop{\hat{\psi}_b(\mathbf{r},t)} \right\rangle
= &   \int \mathop{\mathrm{d}^3 r_1} \ldots \mathop{\mathrm{d}^3 r_N}  \left\langle 
\mathop{\hat{\psi}^{\dag}_a(\mathbf{r},t)} 
\mathop{\hat{\psi}_b(\mathbf{r},t)} \right\rangle_{\{\mathbf{r}_1,\dots,\mathbf{r}_N\}} 
\mathop{P(\mathbf{r}_1,\dots,\mathbf{r}_N)}, \\
\label{eq:sampling2}
\left\langle 
\mathop{\hat{\psi}^{\dag}_a(\mathbf{r},t)} 
\mathop{\hat{\psi}^{\dag}_b(\mathbf{r}',t)} 
\mathop{\hat{\psi}_c(\mathbf{r}',t)} 
\mathop{\hat{\psi}_d(\mathbf{r},t)} \right\rangle 
= &
  \int \mathop{\mathrm{d}^3 r_1} \ldots \mathop{\mathrm{d}^3 r_N} 
\left\langle 
\mathop{\hat{\psi}^{\dag}_a(\mathbf{r},t)} 
\mathop{\hat{\psi}^{\dag}_b(\mathbf{r}',t)} 
\mathop{\hat{\psi}_c(\mathbf{r}',t)} 
\mathop{\hat{\psi}_d(\mathbf{r},t)} \right\rangle_{\{\mathbf{r}_1,\dots,\mathbf{r}_N\}} 
\mathop{P(\mathbf{r}_1,\dots,\mathbf{r}_N)},
\end{align}
\end{widetext}
and so forth for higher-order correlations, where the field operators $\hat{\psi}_{a}^{\dag}(\mathbf{r})$ and $\hat{\psi}_a(\mathbf{r})$ create and annihilate atoms in internal state $a\in\{g,e\}$ at position $\mathbf{r}$.
The atomic correlation functions for a single realization of fixed atomic positions $\{\mathbf{r}_1,\dots,\mathbf{r}_N\}$ (as indicated by the subscript) are given in terms of $\rho_{ab}^{(j)}$ and $\rho_{ad;bc}^{(j,\ell)}$ by
\begin{align} 
\label{eq:pabExp}
\left\langle \mathop{\hat{\psi}^{\dag}_a(\mathbf{r},t)} \mathop{\hat{\psi}_b(\mathbf{r},t)} \right\rangle_{\{\mathbf{r}_1,\dots,\mathbf{r}_N\}}
= 
\sum_j \mathop{\rho_{ab}^{(j)}} (t) &\mathop{\delta(\mathbf{r}-\mathbf{r}_j)} , 
\\ \nonumber
\left\langle \mathop{\hat{\psi}^{\dag}_a(\mathbf{r},t)} \mathop{\hat{\psi}^{\dag}_b(\mathbf{r}',t)} \mathop{\hat{\psi}_c(\mathbf{r}',t)} \mathop{\hat{\psi}_d(\mathbf{r},t)} \right\rangle_{\{\mathbf{r}_1,\dots,\mathbf{r}_N\}}&
\\ \label{eq:padbcExp}
=  \sum_{j\ell} \mathop{\rho_{ad;bc}^{(j,\ell)}}(t)   \mathop{\delta(\mathbf{r} - \mathbf{r}_j)} &\mathop{\delta(\mathbf{r}'-\mathbf{r}_{\ell})} , 
\end{align}
and it is through solving the coupled dynamics between the light and atoms for each stochastic run and ensemble-averaging over many such realizations that we establish the light-induced spatial correlations between atoms~\cite{Javanainen1999,Lee2016}.   

In the limit of low light intensity, the overlap between the incident laser field and the eigenvectors $\mathbf{v}_j$ of the matrix formed by Eq.~\eqref{eq:G_Omegagamma} [ignoring the contact term in Eq.~\eqref{eq:GreensFunction}] determines the resonant behavior of the atomic ensemble \cite{Jenkins2012,JenkinsLongPRB,Bettles2016a}. Because the matrix is complex symmetric rather than Hermitian, the collective eigenmodes $\mathbf{v}_j$ are not necessarily orthogonal, resulting, e.g., in asymmetric Fano-like interference resonances, such as between the in-phase and out-of-phase eigenmodes, as shown in Fig.~2(a) in the main text.

We denote the eigenvalues of  Eq.~\eqref{eq:G_Omegagamma} by $\nu_j+i \upsilon_j$, where $\nu_j= \omega_{eg} - \omega_j$ are the shifts of the collective mode
resonances from the single-atom resonance frequency and $\upsilon_j$ denote the collective radiative half-width at half-maximum (HWHM) linewidths. For $\upsilon_j>\gamma$ ($\upsilon_j<\gamma$ ) the mode is superradiant (subradiant), where $\gamma$ is the independent-atom linewidth. 

For a single, isolated atom, beyond the low light intensity, the solution to \refMasterEquation~(i.e., the optical Bloch equations) in the steady state is
\begin{align} \label{eq:singleAtomOBESolutions}
\pge =& \sqrt{\frac{I}{2\Isat}} \frac{-\Delta\gamma + \mathrm{i}\gamma^2}{\Delta^2 + \gamma^2 (1 + I/\Isat)}, 
\\
\label{eq:singleAtomOBESolutionstoo}
\pee = &  \frac{I}{\Isat}\frac{\gamma^2/2}{\Delta^2 + \gamma^2(1+I/\Isat)},
\end{align}
where the intensity $I = 2\epsilon_0 c|\cbEL^+|^2$, the saturation intensity is given by
$\Isat = 4\pi^2\hbar\gamma c/3\lambda^3$, 
and the linewidth $\gamma$ of the terms in Eq.~(\ref{eq:singleAtomOBESolutions}) and Eq.~(\ref{eq:singleAtomOBESolutionstoo}) experiences a power broadening $\gamma_{\mathrm{PB}} = \gamma\sqrt{1 + I/\Isat}$. 

We compare the full quantum solution of \QME [\refMasterEquation] with the semiclassical equations (\SC) for the single-body terms $\rho_{ab}^{(j)}$ based on the factorization $\rho_{ad;bc}^{(j,\ell)} \simeq \rho_{ad}^{(j)}\rho_{bc}^{(\ell)}$
[Eqs.~(4) and (5) in the main text], which neglects quantum fluctuations. In terms of the stochastic sampling procedure, we express this semiclassical factorization as 
\begin{widetext}
\begin{equation} \label{semicorre}
\left\langle 
\mathop{\hat{\psi}^{\dag}_a(\mathbf{r})} 
\mathop{\hat{\psi}^{\dag}_b(\mathbf{r}')} 
\mathop{\hat{\psi}_c(\mathbf{r}')} 
\mathop{\hat{\psi}_d(\mathbf{r})} \right\rangle_{\rm SC} =   
\int 
\mathop{\mathrm{d}^3 r_1} \ldots
\mathop{\mathrm{d}^3 r_N}
\left\langle 
\mathop{\hat{\psi}^{\dag}_a(\mathbf{r})}
\mathop{\hat{\psi}_d(\mathbf{r})} \right\rangle_{\{\mathbf{r}_1,\dots,\mathbf{r}_N\}}   
\left\langle 
\mathop{\hat{\psi}^{\dag}_b(\mathbf{r}')}
\mathop{\hat{\psi}_c(\mathbf{r}')} \right\rangle_{\{\mathbf{r}_1,\dots,\mathbf{r}_N\}}  
\mathop{P(\mathbf{r}_1,\dots,\mathbf{r}_N)} 
\left[1-\mathop{\delta(\mathbf{r}-\mathbf{r}')}\right],
\end{equation}
\end{widetext}
where the $[1-\delta(\mathbf{r}-\mathbf{r}')]$ term is necessary to exclude the case where the annihilation operators refer twice to the same atom.
Despite the factorization of the internal atomic correlation functions, we generally have 
$\langle 
\mathop{\hat{\psi}^{\dag}_a(\mathbf{r})} 
\mathop{\hat{\psi}^{\dag}_b(\mathbf{r}')} 
\mathop{\hat{\psi}_c(\mathbf{r}')} 
\mathop{\hat{\psi}_d(\mathbf{r})} \rangle_{\rm SC} \neq
\langle 
\mathop{\hat{\psi}^{\dag}_a(\mathbf{r})}
\mathop{\hat{\psi}_d(\mathbf{r})} \rangle  
\langle 
\mathop{\hat{\psi}^{\dag}_b(\mathbf{r}')}
\mathop{\hat{\psi}_c(\mathbf{r}')} \rangle
$,
as the fluctuations of the atomic positions that are included in \SC approach can result in strong light-induced correlations.

In general for the atomic distribution before the light enters the sample we have $P(\mathbf{r}_1,\dots,\mathbf{r}_N)= |\Psi (\mathbf{r}_1,\dots,\mathbf{r}_N)  |^2$, where $\Psi (\mathbf{r}_1,\dots,\mathbf{r}_N) $ denotes the $N$-body atomic wave function in position representation. For the initially uncorrelated atoms, each atom is sampled independently. We consider two different geometries: (i) atoms trapped in a two-dimensional (2D) array with precisely one atom per site; and (ii) a random, uniform distribution of atoms inside a thin, cylindrical disk of radius $R$ and thickness $Z$. 
For the former case, we can sample the stochastic position of an atom in each site~\cite{Jenkins2012}, obtaining $P(\mathbf{r}_1,\dots,\mathbf{r}_N)=\varrho_1(\mathbf{r}_1) \ldots \varrho_N(\mathbf{r}_N)$, where the density distribution of the $j$th array site, centered at $\mathbf{r}_j=\mathbf{R}_j$, is approximated by a Gaussian
\begin{equation}
\varrho_j(\mathbf{r}_j) = 
\frac{\displaystyle\mathrm{exp}\left( -\frac{[x_j-X_j]^2}{2 \sigma_x^2} -\frac{[y_j-Y_j]^2}{2 \sigma_y^2} -\frac{[z_j-Z_j]^2}{2 \sigma_z^2} \right)}{\displaystyle \sqrt{8\pi^3} \sigma_x \sigma_y \sigma_z},
\end{equation}
where the standard deviations $\sigma_x,\sigma_y,\sigma_z$ quantify the spatial confinement of the trapped atoms in all three directions.

\section{Scattered light}
\noindent The total electric field operator $\vecHat{E}{}^{\pm}(\mathbf{r}) =  \cbEL^{\pm}(\mathbf{r})  + \vecHat{E}{d}^{\pm}(\mathbf{r})$  is the sum of the laser field and the fields scattered from all atoms
\begin{align} \label{eq:scatteredFieldDefinition}
\epsilon_0\mathop{\vecHat{E}{d}^+(\mathbf{r})} 
 = \int \mathop{\mathrm{d}^3 \rAtom} 
 \mathop{\mathsf{G}(\mathbf{r}-\rAtomBold)} 
 \mathop{\hat{\mathbf{P}}^+(\rAtomBold)} ,
\end{align}
where $\hat{\mathbf{P}}^+(\rAtomBold) = \mathbf{d}_{ge} \mathop{\hat{\psi}^{\dag}_g (\rAtomBold)} \mathop{\hat{\psi}_e (\rAtomBold)}$ is the atomic polarization.

To analyze the different contributions in the scattered light, we write it as $\vecHat{E}{d}^+ = \langle \vecHat{E}{d}^+ \rangle + \delta \vecHat{E}{d}^+$, where $\delta\vecHat{E}{d}^+$ denotes the fluctuations. 
We then obtain
\begin{equation}
\label{eq:introducingFluctuations}
\begin{split} 
\left\langle 
\mathop{\hat{\mathbf{E}}^-(\mathbf{r})} 
\mathop{\hat{\mathbf{E}}^+(\mathbf{r})} \right\rangle
= & 
\mathop{\cbEL^-(\mathbf{r})} 
\mathop{\cbEL^+(\mathbf{r})} +
\cbEL^-(\mathbf{r}) \left\langle\vecHat{E}{d}^+(\mathbf{r})\right\rangle 
\\&+
\left\langle\vecHat{E}{d}^-(\mathbf{r})\right\rangle \cbEL^+(\mathbf{r}) 
+
\left\langle \vecHat{E}{d}^-(\mathbf{r}) \right\rangle
\left\langle \vecHat{E}{d}^+(\mathbf{r}) \right\rangle 
\\&+
\left\langle \mathop{\delta\vecHat{E}{d}^-(\mathbf{r})} 
\mathop{\delta\vecHat{E}{d}^+(\mathbf{r})}\right\rangle;
\end{split}
\end{equation}
here $\hat{\mathbf{E}}^{-}\hat{\mathbf{E}}^{+}$ is a dyadic product with elements $\hat{E}^-_{\alpha} \hat{E}^+_{\beta}$, with $\alpha,\beta \in\{1,2,3\}$ cycling over the different polarization components, where the intensity is proportional to its diagonal elements.
The first term on the right hand side of Eq.~\eqref{eq:introducingFluctuations} yields the incident light intensity, the second, third, and fourth terms produce the coherent scattering, and the final term produces incoherent scattering dependent on fluctuations. 
Rearranging Eq.~\eqref{eq:introducingFluctuations} to solve for the incoherent scattering gives
\begin{equation} \label{eq:incoherentScatteringFieldExpectation}
\left\langle \delta\vecHat{E}{d}^-(\mathbf{r}) \,\delta\vecHat{E}{d}^+(\mathbf{r}) \right\rangle = 
\left\langle\hat{\mathbf{E}}_d^-(\mathbf{r}) \,\hat{\mathbf{E}}_d^+(\mathbf{r}) \right\rangle - 
\left\langle\hat{\mathbf{E}}_d^-(\mathbf{r})\right\rangle
\left\langle\hat{\mathbf{E}}_d^+(\mathbf{r})\right\rangle,
\end{equation}
which describes correlations in the scattered light.

Consider first a single atom at the origin $\rAtomBold=0$. According to Eq.~\eqref{eq:scatteredFieldDefinition}, the coherently scattered light consists of expectation values
\begin{equation} \label{oneatomsc1}
\left\langle \mathop{\hat{\mathbf{E}}_d^+(\mathbf{r}) }\right\rangle = 
\frac{1}{\epsilon_0} 
\left[ \mathop{\mathsf{G}(\mathbf{r})} \mathbf{d}_{ge} \right] \langle \hat{\sigma}_- \rangle ,
\end{equation}
and there is no difference between the quantum and semiclassical coherent scattering. 
Hence, any difference between the quantum and semiclassical coherent scattering for a many-atom ensemble is due solely to many-body quantum effects.
The incoherent contribution in Eq.~\eqref{eq:incoherentScatteringFieldExpectation} is more subtle, as
\begin{equation} \label{oneatomsc}
\left\langle \mathop{\hat{\mathbf{E}}_d^-(\mathbf{r}) }\mathop{\hat{\mathbf{E}}_d^+(\mathbf{r}) }\right\rangle = 
\frac{1}{\epsilon_0^2} 
\left[ \mathop{\mathsf{G}(\mathbf{r})} \mathbf{d}_{ge}\right]^*
\left[ \mathop{\mathsf{G}(\mathbf{r})} \mathbf{d}_{ge} \right] \left\langle \hat{\sigma}_{+}\hat{\sigma}_{-} \right\rangle 
\end{equation}
means the incoherently scattered light from a single atom yields
\begin{equation} \label{eq:incoherentScatteringSingleAtom}
\left\langle 
\mathop{\delta\vecHat{E}{d}^-(\mathbf{r})} 
\mathop{\delta\vecHat{E}{d}^+(\mathbf{r})} \right\rangle = 
\frac{1}{\epsilon_0^2} 
\left[ \mathop{\mathsf{G}(\mathbf{r})} \mathbf{d}_{ge}\right]^*
\left[ \mathop{\mathsf{G}(\mathbf{r})} \mathbf{d}_{ge} \right]  \left( \langle \hat{\sigma}_{ee} \rangle -| \langle \hat{\sigma}_+\rangle |^2  \right),
\end{equation}
where we have used $\hat{\sigma}_+ \hat{\sigma}_- = \hat{\sigma}_{ee}$.
In the semiclassical approximation, where the quantum fluctuations are ignored, one then replaces $\hat{\sigma}_+$ by $\langle \hat{\sigma}_+\rangle $ in Eq.~\eqref{oneatomsc} \cite{Meystre2007}, such that 
\begin{equation} \label{oneatomsemi}
\left\langle \mathop{\hat{\mathbf{E}}_d^-(\mathbf{r}) }\mathop{\hat{\mathbf{E}}_d^+(\mathbf{r}) }\right\rangle_{\rm SC} = 
\frac{1}{\epsilon_0^2} 
\left[ \mathop{\mathsf{G}(\mathbf{r})} \mathbf{d}_{ge}\right]^*
\left[ \mathop{\mathsf{G}(\mathbf{r})} \mathbf{d}_{ge} \right] \left|\langle \hat{\sigma}_+ \rangle \right|^2,
\end{equation}
and the incoherently scattered light intensity in Eq.~(\ref{eq:incoherentScatteringSingleAtom}) vanishes. Unlike the coherent scattering, the incoherent scattering for a single atom therefore differs depending on whether we treat it in a quantum or semiclassical manner.

Generalizing to the many-atom case, Eq.~(\ref{oneatomsc}) now becomes 
\begin{widetext}
\begin{equation} \label{eq:EdEdExpectation}
\left\langle \mathop{\hat{\mathbf{E}}_d^-(\mathbf{r}) }\mathop{\hat{\mathbf{E}}_d^+(\mathbf{r}) }\right\rangle 
= 
\frac{1}{\epsilon_0^2} \int \mathop{\mathrm{d}^3 \rAtom}
\mathop{\mathrm{d}^3 \rAtom'}
\left[ \mathsf{G}(\mathbf{r}-\rAtomBold) \right]^*
\left[ \mathsf{G}(\mathbf{r}-\rAtomBold') \right] 
\left\langle \mathop{\hat{\mathbf{P}}^-(\rAtomBold)} \mathop{\hat{\mathbf{P}}^+(\rAtomBold')}  \right\rangle,
\end{equation}
\end{widetext}
where, as in Eq.~\eqref{eq:scatteredFieldDefinition}, $[\mathsf{G}(\mathbf{r}-\rAtomBold)]^*$ acts on $\hat{\mathbf{P}}^-(\rAtomBold)$ and likewise $\mathsf{G}(\mathbf{r}-\rAtomBold')$ on $\hat{\mathbf{P}}^+(\rAtomBold')$. When calculating the full quantum solution, the correlation functions are evaluated using the solution to \QME [\refMasterEquation] and by ensemble-averaging over many realizations of atomic positions. However, we can also introduce the many-body version of the single-atom semiclassical approximation [Eq.~\eqref{oneatomsemi}] to light scattering:
\begin{widetext}
\begin{equation} \label{semiresponse}
\left\langle \hat{\mathbf{P}}^-(\rAtomBold) \hat{\mathbf{P}}^+(\rAtomBold')  \right\rangle \simeq 
 \int \mathop{\mathrm{d}^3 r_1} \ldots \mathop{\mathrm{d}^3 r_N} 
 \left\langle \hat{\mathbf{P}}^-(\rAtomBold) \right\rangle_{\rjrN}   \left\langle \hat{\mathbf{P}}^+(\rAtomBold') \right\rangle_{\rjrN}  
 \mathop{P(\mathbf{r}_1,\dots,\mathbf{r}_N)},
\end{equation}
substituting this back into Eq.~\eqref{eq:EdEdExpectation} to give the semiclassical scattered field  
\begin{equation} \label{eq:scatteredFieldExpectationSC}
\left\langle \mathop{\hat{\mathbf{E}}_d^-(\mathbf{r}) }\mathop{\hat{\mathbf{E}}_d^+(\mathbf{r}) }\right\rangle_{\mathrm{SC}} 
= 
\frac{1}{\epsilon_0^2}  \int \mathop{\mathrm{d}^3 \rAtom} 
\mathop{\mathrm{d}^3 \rAtom'}
\left[ \mathsf{G}(\mathbf{r}-\rAtomBold) \right]^*
\left[ \mathsf{G}(\mathbf{r}-\rAtomBold') \right]
\int \mathop{\mathrm{d}^3 r_1} \ldots \mathop{\mathrm{d}^3 r_N} 
 \left\langle \hat{\mathbf{P}}^-(\rAtomBold) \right\rangle_{\rjrN}   \left\langle \hat{\mathbf{P}}^+(\rAtomBold') \right\rangle_{\rjrN}  
 \mathop{P(\mathbf{r}_1,\dots,\mathbf{r}_N)}.
\end{equation}
\end{widetext}

Deriving the semiclassical scattered light in Eq.~\eqref{eq:scatteredFieldExpectationSC} corresponds to a systematic way of neglecting all quantum fluctuations when the atomic response is first calculated from \SC [Eqs.~(4) and (5)]. Hence, comparing the scattered light of Eq.~\eqref{eq:scatteredFieldExpectationSC} with the equivalent full quantum solution of Eq.~\eqref{eq:EdEdExpectation} provides a signature for quantum effects in the collective atomic response. Alternatively, if our goal is to determine a computationally efficient and accurate approximation to the full quantum solution, we can instead try to improve the semiclassical approximation. A simple way to achieve this without increasing computational demands is to include the single-atom quantum description to incoherent scattering [Eq.~\eqref{eq:incoherentScatteringSingleAtom}] integrated over the extent of the sample, which is sufficient in a number of cases to capture the leading quantum contributions.

We begin this procedure by placing the atomic operators in Eq.~\eqref{eq:EdEdExpectation} in the normal order. This yields for the expectation term  on the right hand side of Eq.~\eqref{eq:EdEdExpectation} (for both fermionic and bosonic atoms)
\begin{multline} 
\label{eq:normalOrder}
\left\langle  
\mathop{\hat{\psi}_e^{\dag}(\rAtomBold)}
\mathop{\hat{\psi}_g(\rAtomBold)}
\mathop{\hat{\psi}_g^{\dag}(\rAtomBold')} 
\mathop{\hat{\psi}_e(\rAtomBold')} \right\rangle
= 
\left\langle  \mathop{\hat{\psi}_e^{\dag}(\rAtomBold)}
\mathop{\hat{\psi}_e(\rAtomBold')}\right\rangle 
\mathop{\delta(\rAtomBold-\rAtomBold')} 
\\
+ 
\left\langle  
\mathop{\hat{\psi}_e^{\dag}(\rAtomBold)} 
\mathop{\hat{\psi}_g^{\dag}(\rAtomBold')} 
\mathop{\hat{\psi}_e(\rAtomBold')} 
\mathop{\hat{\psi}_g(\rAtomBold)} \right\rangle.
\end{multline}
Substituting this into Eq.~\eqref{eq:EdEdExpectation} and using the semiclassical factorization approximation of Eq.~\eqref{semicorre} we obtain
\begin{widetext}
\begin{multline} 
\label{eq:scatteredFieldNormalOrderSAQ}
\left\langle \mathop{\hat{\mathbf{E}}_d^-(\mathbf{r}) }\mathop{\hat{\mathbf{E}}_d^+(\mathbf{r}) }\right\rangle_{\mathrm{SAQ}} 
= 
\frac{1}{\epsilon_0^2} \int \mathop{\mathrm{d}^3 \rAtom}
\left[ \mathop{\mathsf{G}(\mathbf{r}-\rAtomBold)} \mathbf{d}_{ge}\right]^{*}
\left[ \mathop{\mathsf{G}(\mathbf{r}-\rAtomBold)} \mathbf{d}_{ge} \right]
\left\langle  \mathop{\hat{\psi}_e^{\dag}(\rAtomBold)}  
\mathop{\hat{\psi}_e(\rAtomBold)}\right\rangle   
\\
+  \frac{1}{\epsilon_0^2} \int' \mathop{\mathrm{d}^3 \rAtom} 
\mathop{\mathrm{d}^3 \rAtom'} \left\{
\left[ \mathsf{G}(\mathbf{r}-\rAtomBold) \right]^*
\left[ \mathsf{G}(\mathbf{r}-\rAtomBold') \right]
 \int \mathop{\mathrm{d}^3 r_1} \ldots \mathop{\mathrm{d}^3 r_N}   
 \left\langle \hat{\mathbf{P}}^-(\rAtomBold) \right\rangle_{\{\mathbf{r}_1,\dots,\mathbf{r}_N\}}   \left\langle \hat{\mathbf{P}}^+(\rAtomBold') \right\rangle_{\{\mathbf{r}_1,\dots,\mathbf{r}_N\}}  
 \mathop{P(\mathbf{r}_1,\dots,\mathbf{r}_N)}  \right\},
\end{multline}
where $\int'$ denotes a double integral over all $\{\rAtom,\rAtom'\}$ excluding $\rAtom=\rAtom'$.
The difference between this augmented (semiclassical plus single-atom quantum) expression and the semiclassical expression of Eq.~\eqref{eq:scatteredFieldExpectationSC} in the scattered intensity is effectively the contributions of the single atom incoherent (quantum) scattering from Eq.~\eqref{eq:incoherentScatteringSingleAtom} integrated over the extent of the sample:
\begin{multline} 
\label{eq:SAQSC}
\left\langle \mathop{\hat{\mathbf{E}}_d^-(\mathbf{r}) }\mathop{\hat{\mathbf{E}}_d^+(\mathbf{r}) } \right\rangle_{\mathrm{SAQ}}    -  \left\langle \mathop{\hat{\mathbf{E}}_d^-(\mathbf{r}) }\mathop{\hat{\mathbf{E}}_d^+(\mathbf{r}) }\right\rangle_{\mathrm{SC}} 
 = 
 \frac{1}{\epsilon_0^2} \int \mathop{\mathrm{d}^3 \rAtom}
\left[ \mathop{\mathsf{G}(\mathbf{r}-\rAtomBold)} \mathbf{d}_{ge}\right]^{*}
\left[ \mathop{\mathsf{G}(\mathbf{r}-\rAtomBold)} \mathbf{d}_{ge} \right]
\left\langle  \mathop{\hat{\psi}_e^{\dag}(\rAtomBold)}
\mathop{\hat{\psi}_e(\rAtomBold)}\right\rangle   
\\
-
\frac{1}{\epsilon_0^2} \int \mathop{\mathrm{d}^3 \rAtom}
\left\{
\left[ \mathsf{G}(\mathbf{r}-\rAtomBold) \right]^{*}
\left[ \mathsf{G}(\mathbf{r}-\rAtomBold) \right]
 \int \mathop{\mathrm{d}^3 r_1} \ldots \mathop{\mathrm{d}^3 r_N} 
 \left\langle \hat{\mathbf{P}}^-(\rAtomBold) \right\rangle_{\{\mathbf{r}_1,\dots,\mathbf{r}_N\}}   \left\langle \hat{\mathbf{P}}^+(\rAtomBold) \right\rangle_{\{\mathbf{r}_1,\dots,\mathbf{r}_N\}}  
\mathop{P(\mathbf{r}_1,\dots,\mathbf{r}_N)}  \right\}.
\end{multline}
\end{widetext}
This improved description includes both the semiclassical contribution and the single-body quantum fluctuations, meaning any difference in the incoherent scattering between this improved model and the full quantum model is solely due to many-body quantum effects.

\section{Transmission}

\noindent
\noindent
In this work we consider coherently and incoherently transmitted light and calculate them through a disk of cross-sectional area $S$ perpendicular to the optical axis a distance $f=500\lambda$ downstream of the atoms. We consider light transmitted in the same spatial mode as the driving field, motivated by a typical experimental scheme of collecting transmitted light into a single-mode optical fibre, although, for simplicity, we ignore any explicit refocussing or fibre coupling.  The transmitted light therefore has the form 
\begin{equation} 
\label{eq:totalTransmissionDefinition}
T = \frac{\displaystyle \int \mathop{\mathrm{d}S}\int \mathop{\mathrm{d}S'} \cbEL^+(\mathbf{r}) \cdot \left\langle \mathop{\hat{\mathbf{E}}^-(\mathbf{r})} \mathop{\hat{\mathbf{E}}^+(\mathbf{r}')}\right\rangle \cdot \mathop{\cbEL^-(\mathbf{r}')}}
{\displaystyle \left|\int \mathop{\mathrm{d}S} \mathop{\cbEL^+(\mathbf{r})} \cdot \mathop{\cbEL^-(\mathbf{r})} \right|^2 }.
\end{equation}
Note that because of the double integral over $S$ and $S'$
the expectation term is now a function of $\mathbf{r}$ and $\mathbf{r}'$, although substituting $\mathbf{r}'$ into the preceding equations does not affect our discussion of coherent and incoherent scattering.

To calculate the coherent transmission $T_{\mathrm{coh}}$ (plotted as optical depth $\mathrm{OD}\equiv -\log T_{\mathrm{coh}}$), we substitute the first four terms on the right hand side of Eq.~\eqref{eq:introducingFluctuations}  into Eq.~\eqref{eq:totalTransmissionDefinition}. This gives quantum  $T_{\mathrm{coh}}^{\mathrm{QM}}$ or semiclassical  $T_{\mathrm{coh}}^{\mathrm{SC}}$ coherent transmission, depending on whether we use the solutions to  \QME or \SC. To calculate the incoherent contribution to the transmission, we replace the two-field expectation in Eq.~\eqref{eq:totalTransmissionDefinition} with Eq.~\eqref{eq:incoherentScatteringFieldExpectation}. Evaluating Eq.~\eqref{eq:incoherentScatteringFieldExpectation} using Eqs.~\eqref{eq:EdEdExpectation} and \eqref{eq:normalOrder}, along with the solutions to \QME, results in the quantum incoherent transmission $T_{\mathrm{inc}}^{\mathrm{QM}}$. Using instead the solutions to \SC and either Eq.~\eqref{eq:scatteredFieldExpectationSC} or Eq.~\eqref{eq:scatteredFieldNormalOrderSAQ}, respectively, produces the semiclassical incoherent transmission $T_{\mathrm{inc}}^{\mathrm{SC}}$, or the improved model for incoherent transmission $T_{\mathrm{inc}}^{\mathrm{SAQ}}$ where the independent-atom quantum description is added to the semiclassical model.
\\

%% figure %%%
\begin{figure}[tb]
\centering
\includegraphics[width=8.4cm]{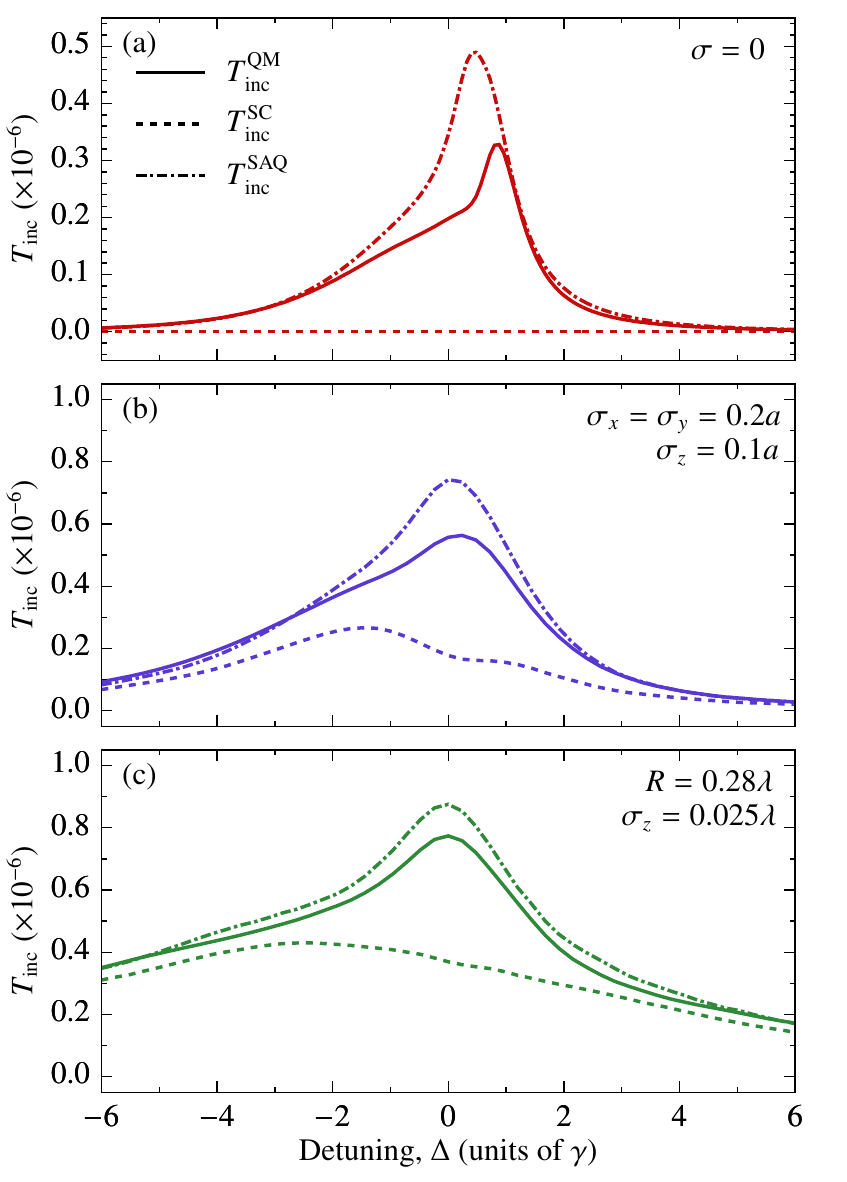}\vspace{-0.1cm}
\caption{Different quantum and classical sources of fluctuations on the incoherent forward scattering. The parameters are as in Fig.~2 of the main text, except the intensity $I=\Isat$. A $2\times 2$ square lattice for (a) fixed and (b) fluctuating ($\sigma_{x,y}=0.2a, \sigma_z=0.025\lambda$) atomic positions, and (c) a random uniformly-distributed disk of atoms with radius $R=0.28\lambda$ and $\sigma_z=0.025\lambda$. Dashed lines correspond to the semiclassical results $T_{\mathrm{inc}}^{\mathrm{SC}}$ where the only source of fluctuations are classical spatial fluctuations. The additional effect of single-atom quantum fluctuations is included in the dot-dashed lines, $T_{\mathrm{inc}}^{\mathrm{SAQ}}$. Solid lines are obtained by the full quantum treatment $T_{\mathrm{inc}}^{\mathrm{QM}}$, also accounting for the many-body quantum fluctuations.}
\label{fig:ComparingIncoherentScatteringModels}
\end{figure}

\section{Fluctuations in incoherent light}

\noindent In Figs.~2 and 3 of the main section we analyzed the incoherent transmission and found it to be well approximated by $T_{\mathrm{inc}}^{\mathrm{SAQ}}$, with $T_{\mathrm{inc}}^{\mathrm{SC}}$ providing a negligible contribution at high intensities.
In Fig.~\ref{fig:ComparingIncoherentScatteringModels} we show a special case with $I=\Isat$ where all the different sources of fluctuations can have comparable contributions. We consider (a) fixed and (b) fluctuating positions, and (c) in a random uniformly distributed 2D disk. In the absence of position fluctuations (a), while there are no classical fluctuations in the incoherent scattering ($T_{\mathrm{inc}}^{\mathrm{SC}}$), close to resonance both the single-atom quantum description of light emission ($T_{\mathrm{inc}}^{\mathrm{SAQ}}-T_{\mathrm{inc}}^{\mathrm{SC}}$) and many-body quantum fluctuations ($T_{\mathrm{inc}}^{\mathrm{QM}} - T_{\mathrm{inc}}^{\mathrm{SAQ}}$) are equally important. Adding in position fluctuations (b) results in classical fluctuations such that all three contributions are significant in the overall signal, and for this choice of density even the completely random disk (c) still exhibits many-body quantum fluctuations.

%% figure %%
\begin{figure*}[tb]
\centering
\includegraphics[width=\textwidth]{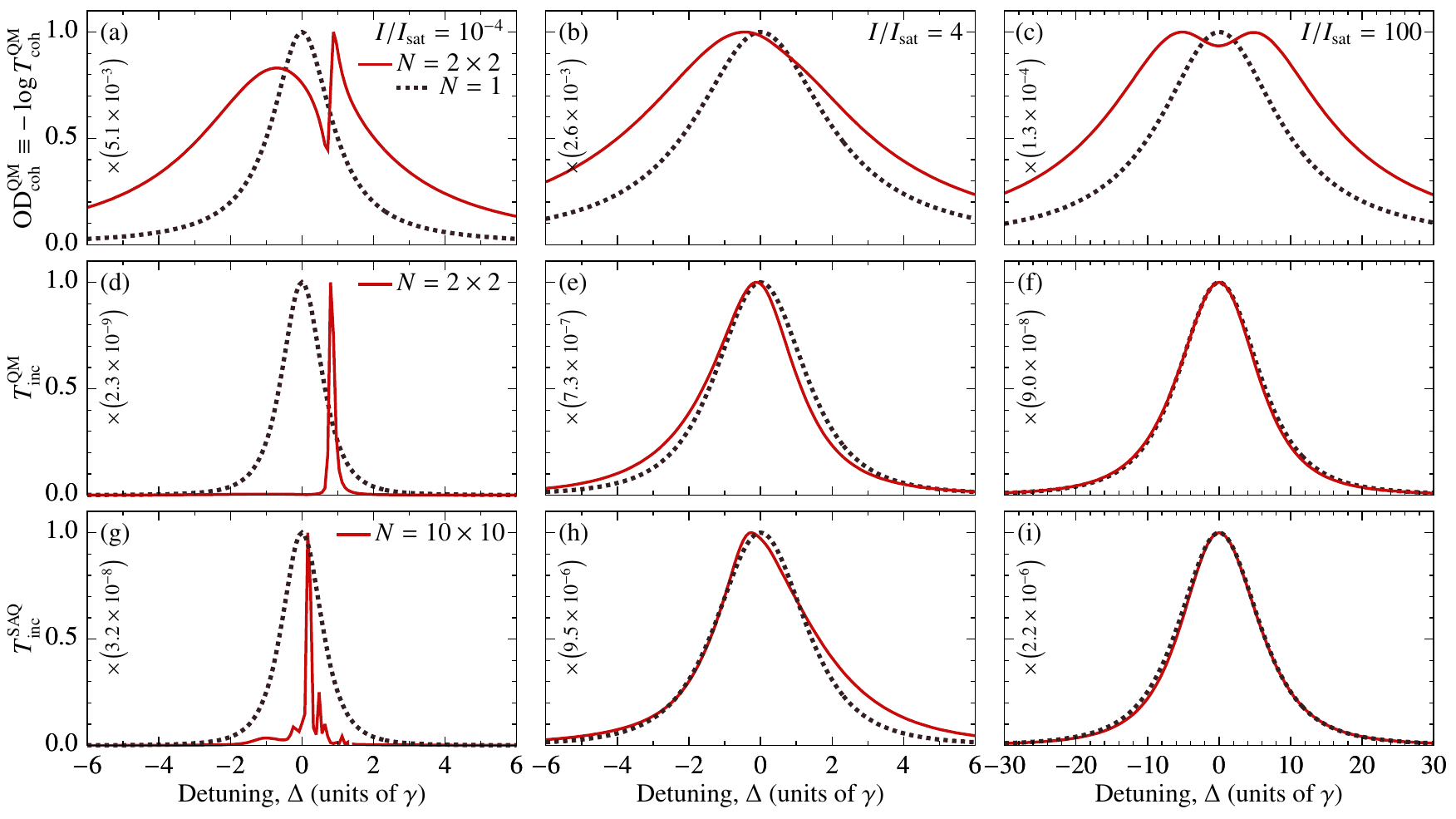}\vspace{-0.1cm}
\caption{Collective effects (red solid lines)  in transmitted light compared with a single atom fixed at the focus (black dotted lines).
(a--c) The optical depth of the coherent transmission calculated with \QME, $\mathrm{OD}_{\mathrm{coh}}^{\mathrm{QM}}=-\log T_{\mathrm{coh}}^{\mathrm{QM}}$ and the incoherent transmission calculated using (d--f)  \QME $T_{\mathrm{inc}}^{\mathrm{QM}}$ and (g--i)  the semiclassical model with single-atom quantum description $T_{\mathrm{inc}}^{\mathrm{SAQ}}$.
We consider a $2\times 2$ (a--f) and $10\times 10$ (g--i) array with fixed positions, using parameters, respectively, from Figs.~2 and 6(a,b) in the main text. 
Black dotted lines are normalized to the same amplitude as the peak of the red solid lines. The laser intensity is (a,d) $I/\Isat = 10^{-4}$, (b,e) $I/\Isat=4$, and (c,f) $I/\Isat=100$. }
\label{fig:SingleAtom}
\end{figure*}

\section{Scattering from a single atom}

\noindent To illustrate how many-atom collective response affects transmitted light, we show in Fig.~\ref{fig:SingleAtom} the coherent and incoherent transmission for a $2\times 2$ array, incoherent transmission for a $10\times 10$ array, 
and for a single atom. 
For a single atom, the coherent scattering (a--c) is purely classical since $\TcohQM=\TcohSC$ [Eq.~\eqref{oneatomsc1}], while the incoherent scattering (d--f) is purely quantum since from Eqs.~\eqref{eq:incoherentScatteringSingleAtom} 
and \eqref{oneatomsemi} we get $\TincQM\neq(\TincSC=0)$. For the $2\times 2$ array we plot $\TcohQM$ (a--c) and $\TincQM$ (d--f); compare these with $\TcohSC$ and $\TincSAQ$ of Fig.~2 that use the same parameters as
Fig.~\ref{fig:SingleAtom}(a--f)

For the coherent transmission [Fig.~\ref{fig:SingleAtom}(a--c)], the single atom lineshape is given by a single resonance with power-broadened linewidth $\gamma_{\mathrm{PB}}=\gamma\sqrt{1+I/\Isat}$ [Eq.~\eqref{eq:singleAtomOBESolutions}]. 
The many-atom lineshapes, however, exhibit clear qualitative differences, including multiple resonances and modified power-broadened linewidths, clearly indicating the effects of the 
sample geometry and light-mediated interactions.

The incoherent transmission through a  $10\times 10$ array  [Fig.~\ref{fig:SingleAtom}(g--i)]  is calculated using the semiclassical model incorporating the single-atom quantum description, $T_{\mathrm{inc}}^{\mathrm{SAQ}}$, using the same parameters as the coherent transmission in Fig.~6(a,b) of the main text. Similarly to the full quantum $2\times 2$ case in (d--f), the lineshape approaches the single atom quantum lineshape with HWHM $5.5\gamma$.

\section{Incident laser field}
\subsection{Paraxial Gaussian beam}
\noindent The amplitude of a paraxial Gaussian laser beam in the absence of atoms propagating in the $z$ direction and focused at $z=0$ has the form 
\begin{equation} \label{eq:GaussianBeam}
\cbEL^+(\mathbf{r}) = \cEL^+ \frac{w_0}{w} \mathrm{e}^{\mathrm{i}kz} \mathrm{e}^{\mathrm{i}k\rho_z^2/2R_c} \mathrm{e}^{-\mathrm{i}\zeta(z)} \mathrm{e}^{-\rho_z^2/w^2} \polvec ,
\end{equation}
where $\cEL^+$ is the maximum amplitude, $w=w_0 (1+z^2/\zR^2)^{1/2}$ the beam radius, $w_0$ the beam waist, $\zR=\pi w_0^2/\lambda$ the Rayleigh range, $\rho_z= (x^2+y^2)^{1/2}$, $R_c=z+\zR^2/z$ the beam curvature, $\zeta(z)=\mathrm{arctan}(z/\zR)$ the Gouy phase, and $\polvec$ the unit polarization vector. In every figure except for 
Fig.~6(c), 
the beam waist $w_0=10\lambda$ is sufficiently large for the paraxial model to be a good approximation to the true beam propagation.

\subsection{Vector Gaussian beam}
\noindent In Fig.~6(c) 
of the main text we consider beam waists of $w_0\leq 3\lambda$, at which point the vector nature of 
the light must be correctly accounted for. This is carried out numerically with the method used in \cite{Bettles2016,Tey2009,Adams2018}.
We consider a field $\cEL^{+}\mathrm{e}^{-\rho_L^2/w_L^2}\polvec_+$ with a Gaussian profile incident on a lens at position $z=-f$, where $f$ is the focal length of the lens, $\rho_L$ and $w_L$ are respectively the radial position and beam radius at the lens, and $\polvec_{\pm} = \mp(\hat{\mathbf{x}}\pm\mathrm{i}\hat{\mathbf{y}})/\sqrt{2}$ is a circular polarization unit vector. Immediately after passing through the (ideal) lens, the field has the form
\begin{widetext}
\begin{equation} \label{eq:ELvec}
\cbEL^+ (\rho_L, \phi, z=-f) = \frac{\cEL^{+} \mathop{\mathrm{e}^{-\rho_L^2/w_L^2}} }{\sqrt{|\cos\theta|}} \left( \frac{1 + \cos \theta}{2} \polvec_+ + \frac{\sin\theta}{\sqrt{2}} \mathrm{e}^{\mathrm{i}\phi} \hat{\mathbf{z}} + \frac{ \cos \theta -1}{2} \mathrm{e}^{2\mathrm{i}\phi} \polvec_- \right)
\mathop{\exp\left(-\mathrm{i} \left[k \sqrt{\rho_L^2 + f^2} - \pi/2 \right] \right)},
\end{equation}
\end{widetext}
where $\phi = \tan^{-1}(y/x)$, and $\theta=\tan^{-1}(\rho_L/f)$ is the angle between the $-z$ axis and a point on the lens. To calculate the field propagation, it is helpful to decompose the field into an orthogonal set of modes: $\cbEL^+ = \sum_{\mu} \kappa_{\mu} \cE_{\mu}^+$, where $\mu = (k_t,s,m)$, $k_t=\sqrt{k^2-k_z^2}$ is the transverse wavevector component, $s=\pm1$ is the helicity, $m$ is an angular momentum index, and the expansion coefficients are given by
\begin{widetext}
\begin{multline} 
\label{eq:kappa_mu}
\kappa_{\mu} = \delta_{m1} \pi k_t \int_0^{\infty} \mathop{\mathrm{d}\rho_L}  \frac{\rho_L}{\sqrt{\cos\theta}} \Bigg\{\frac{sk+k_z}{k} \left( \frac{1+\cos\theta}{2} \right) 
\mathop{J_0(k_{t}  \rho_L)} + \mathrm{i}\frac{\sqrt{2}k_t}{k} \left( \frac{\sin\theta}{\sqrt{2}} \right) \mathop{J_1(k_{t} \rho_L)} 
\\ + \frac{sk-k_z}{k} \left( \frac{\cos\theta-1}{2} \right) \mathop{J_2(k_t \rho_L)} \Bigg\} \exp \left( -\mathrm{i} \left[ k\sqrt{\rho_L^2 + f^2} - \pi/2 \right] - \frac{\rho_L^2}{w_L^2} \right),
\end{multline}
\end{widetext}
with the $J_m$ describing $m$th order Bessel functions. The field, taken at a distance $z$ from the lens focus 
(located at the origin), 
is then given in terms of the decomposed $\sigma_\pm$ and $z$ polarization components by 
\pagebreak

\noindent\begin{align} 
\label{eq:FullVecFieldPropagation}
\cE_+^+ (\rho_z,\phi,z) =&  \cEL^+ \sum_{s=\pm1} \int^k_0 \mathop{\mathrm{d}k_t}  \frac{s k + k_z}{4\pi k} \mathop{J_0(k_t\rho_z)} \mathrm{e}^{\mathrm{i}k_z (z+f)} \kappa_{\mu} , 
\\
\label{eq:FullVecFieldPropagation3}
\cE_-^+ (\rho_z,\phi,z) =&  \cEL^+ \sum_{s=\pm1} \int^k_0 \mathop{\mathrm{d}k_t}  \frac{s k - k_z}{4\pi k} \mathop{J_2(k_t\rho_z)} \mathop{\mathrm{e}^{\mathrm{i}k_z (z+f)}} \mathop{\mathrm{e}^{2\mathrm{i}\phi}} \kappa_{\mu},
\end{align}
\begin{align}
\label{eq:FullVecFieldPropagation2}
\cE_z^+ (\rho_z,\phi,z) = & -\mathrm{i}\cEL^+ \sum_{s=\pm1} \int^k_0 \mathop{\mathrm{d}k_t} \frac{\sqrt{2}k_t}{4\pi k} \mathop{J_1(k_t\rho_z)} \mathrm{e}^{\mathrm{i}k_z (z+f)} \mathop{\mathrm{e}^{\mathrm{i}\phi}} \kappa_{\mu}.
\end{align}
Using Eqs.~(\ref{eq:FullVecFieldPropagation}), (\ref{eq:FullVecFieldPropagation2}) and (\ref{eq:FullVecFieldPropagation3}), we can calculate the total field ${\cbEL^+ = \cE_+^+ \polvec_+ + \cE_-^+ \polvec_- + \cE_z^+ \polvec_z}$ at the location of each individual atom within for example a particular stochastic realization as well as in the output plane at $z=f$.

%% bibliography %%
\bibliography{References_arxiv}